\documentclass[a4paper, amsfonts, amssymb, amsmath, reprint, showkeys, nofootinbib, twoside, superscriptaddress]{revtex4-1}
\usepackage[english]{babel}
\usepackage[utf8]{inputenc}
\usepackage[colorinlistoftodos, color=green!40, prependcaption]{todonotes}
\usepackage{amsthm}
\usepackage{amsmath}
\usepackage{mathtools}
\usepackage{physics}
\usepackage{xcolor}
\usepackage{graphicx}
\usepackage[left=23mm,right=13mm,top=35mm,columnsep=15pt]{geometry} 
\usepackage{adjustbox}
\usepackage{placeins}
\usepackage[T1]{fontenc}
\usepackage{lipsum}
\usepackage{csquotes}
\usepackage{siunitx}
\usepackage{caption}
\usepackage{subcaption}
\usepackage{ragged2e} 

\usepackage[pdftitle={Article}, pdfauthor={Author}, colorlinks=true]{hyperref}

\begin{document}
\title{Decoherence of a matter-wave interferometer due to dipole-dipole interactions}
\author{Paolo Fragolino}
    \affiliation{Van Swinderen Institute for Particle Physics and Gravity, University of Groningen, 9747 AG Groningen, the Netherlands }
    \affiliation{Dipartimento di Fisica e Astronomia,
Università di Bologna, via Irnerio 46, I-40126 Bologna, Italy}

\author{Martine Schut}
    \affiliation{Van Swinderen Institute for Particle Physics and Gravity, University of Groningen, 9747 AG Groningen, the Netherlands }
    \affiliation{Bernoulli Institute for Mathematics, Computer Science and Artificial Intelligence, University of Groningen, 9747 AG Groningen, the Netherlands \vspace{1mm}}
\author{Marko Toro\v{s}}
    \affiliation{School of Physics and Astronomy, University of Glasgow, Glasgow, G12 8QQ, UK}
\author{Sougato Bose}
    \affiliation{Department of Physics and Astronomy, University College London, Gower Street, WC1E 6BT London, UK}
\author{Anupam  Mazumdar}
    \affiliation{Van Swinderen Institute for Particle Physics and Gravity, University of Groningen, 9747 AG Groningen, the Netherlands }

\begin{abstract}
Matter-wave interferometry with nanoparticles will enable the development of quantum sensors capable of probing ultraweak fields with unprecedented applications for fundamental physics. The high sensitivity of such devices however makes them susceptible to a number of noise and decoherence sources and as such can only operate when sufficient isolation from the environment is achieved. It is thus imperative to model and characterize the interaction of nanoparticles with the environment and to estimate its deleterious effects. The aim of this paper will be to study the decoherence of the matter-wave interferometer due to dipole-dipole interactions which is one of the unavoidable channels for decoherence even for a neutral micro-crystal. We will start the analysis from QED and show that it reduces to the scattering model characterized by the differential cross-section. We will then obtain simple expressions for the decoherence rate in the short and long wavelength limits that can be readily applied to estimate the available coherence time. We will conclude by applying the obtained formulae to estimate the dipole-dipole decoherence rate for the Quantum Gravity-induced Entanglement of Masses (QGEM) protocol and discuss if the effects should be mitigated.
\end{abstract}



\maketitle

\section{Introduction}\label{sec:intro}

The idea that matter can behave as a wave is a key conceptual leap of modern physics~\cite{de1923waves}, with matterwave interferometry one of the central experimental techniques of quantum mechanics~\cite{thomson1927diffraction,davisson1927diffraction,davisson1928reflection}. It is the basis for the notion of quantum superposition~\cite{Schrodinger:1935zz} and it is the building block of quantum entanglement~\cite{einstein1935can,bell1964einstein}, two features that cannot be mimicked by a classical theory~\cite{Horodecki:2009zz}. Matter-wave interferometry has been also used in a series of fundamental experiments to demonstrate gravitationally-induced interference with neutrons and atoms ~\cite{colella1975observation,nesvizhevsky2002quantum,fixler2007atom,asenbaum2017phase,overstreet2022observation}. Furthermore, matter-wave interferometers can be excellent quantum sensors~\cite{Kilian:2022kgm,Wu:2022rdv} and can act as probes of physics beyond the standard model~\cite{Barker:2022mdz}.

 It was further suggested that the next generation of matter-wave interferometers with nanoparticles will be sensitive enough to probe gravitationally-induced entanglement. Known as a quantum gravity-induced entanglement of masses (QGEM)~\cite{Bose:2017nin}~\footnote{The first report of the results of the QGEM protocol \cite{Bose:2017nin} was in a conference in 2016, Bangalore workshop \cite{ICTS}. See also Ref.~\cite{Marletto:2017kzi}.}, the scheme shows that if gravity is inherently a quantum entity then the masses of two nearby interferometers will entangle when placed sufficiently close. The key observation is that as long as we follow the standard relativistic quantum mechanics, locality/causality, and general relativity in an effective field theory of quantum gravity the two quantum superposed masses will inevitably entangle each other via the quantum gravitational interaction~\cite{Bose:2017nin,Marshman:2019sne,Bose:2022uxe,Gunnink:2022ner,Elahi:2023ozf,Vinckers:2023grv,Carney_2019,Belenchia:2018szb,Carney:2021vvt,Danielson:2021egj,Christodoulou:2022vte}, while classical gravity cannot entangle the two quantum systems as formalized by the local operation and classical communication (LOCC) theorem ~\cite{Bennett:1996gf,Bose:2017nin,Marshman:2019sne,Bose:2022uxe}.  Recently, the QGEM protocol was also extended to test the quantum nature of gravity in an optomechanical setup where we can test the quantum gravitational entanglement between matter and photon~\cite{Biswas:2022qto}. However, there are many experimental challenges to be resolved before interferometry with nanoparticles can be implemented. To name a few, creating spatial quantum superpositions~\cite{Bose:2017nin,vandeKamp:2020rqh,Bose:2017nin, PhysRevLett.111.180403,PhysRevLett.117.143003,Margalit:2020qcy,Marshman:2021wyk,Folman2013,folman2019,PhysRevLett.125.023602,Folman2018,Zhou:2022frl,Zhou:2022jug,Zhou:2022epb,Marshman:2023nkh}, ensuring sufficiently long coherence times~\cite{Bose:2017nin,Schut:2021svd,Tilly:2021qef,PhysRevLett.125.023602,Rijavec:2020qxd,RomeroIsart2011LargeQS,Chang_2009,Schlosshauer:2014pgr,bassi2013models,Toros:2020krn}, and protecting the experiment from external jitters, gravity gradient noise and seismic noise~\cite{Toros:2020dbf}. 
 
  The aim of this paper will be to investigate dipole-dipole decoherence in matter-wave interferometry with nanoparticles. Such channel of decoherence is unavoidable and must be taken into account even with neutral micrcrystals~\cite{Casimir:1947kzi,Casimir:1948dh}. 
In section \ref{sec:step2_deco}, we set up the formalism for describing the evolution of the density matrix of a matter-wave interferometer and the environment which are coupled via a generic EM interaction. 
We do this by first constructing the interaction Hamiltonian for QED (subsection~\ref{subsec:hamiltonian}) and then we use this Hamiltonian to find the Born-Markov master equation (subsection~\ref{subsec:master_eqn}).
In section \ref{DD}, we will discuss the dipole-dipole interaction and study the decoherence of the matter-wave interferometer in a short and long wavelength limit (subsections~\ref{sec_short}, \ref{sec_long} respectively).
Section~\ref{sec:QGEM} discusses the application of the decoherence model from dipole interactions to the specific case of the QGEM experiment.
There we discuss all possible dipole-dipole interactions that play a role in the QGEM experiment, such as: the interaction of an induced dipole in the QGEM test mass with an environmental dipole (subsection~\ref{subsec_induced_crystal}), the interaction between a permanent dipole in the test mass with an environmental dipole (subsection~\ref{subsec_d1_d2}), and the interaction between the permanent dipole of the test mass with the induced dipole in the environmental particles that it induces (subsection~\ref{subsec_induced_environment}).
From the three cases we put constraints on the allowed dipoles in the QGEM experiment, which will be discussed in section~\ref{sec:conclusion}.

\section{Modelling decoherence from electromagnetic interactions}\label{sec:step2_deco}

We start by modelling the decoherence caused by a generic electromagnetic interaction between an object in superposition (the matter-wave interferometer)
 and an environmental particle. We will consider a charged neutral crystal which is responsible for creating the spatial quantum superposition for the matter-wave interferometer as a fermion. We will also treat the environmental particle as a fermion field. Hence the two fermions are interacting within the Standard Model by virtue of the photon field. This is the model we will use to study the effect of decoherence in the Born-Markov approximation \cite{Schlosshauer:2014pgr,joos2003decoherence}~\footnote{In this section, we will use the natural units $\hbar=c=1$. We will switch to SI units in section~\ref{sec:QGEM} when we discuss the application to the QGEM experiment.}.

Let us first consider a generic QED interaction Lagrangian between the two fermions. The fermion field $\psi$ that represents the environmental particle will interact electromagnetically via a photon field $A_\mu$.
The fermion-photon interaction term (the three-vertex) is given by:
%
%
%
\begin{equation}
    \mathcal{L}_\text{int}=-e \Bar{\psi}\gamma^{\mu}\psi A_{\mu},
    \label{EM lagrangian}
\end{equation}
%
where $\gamma^{\mu}$ are the Dirac's matrices and $e$ is the electric charge unit. 
Here $\mu=0,1,2,3$, and we take the signature $(-,+,+,+)$ of the metric $\eta_{\mu\nu}$. The corresponding Hamiltonian interaction is given by:
\begin{equation}
    H_\text{int}=- \int d^{3}\textbf{x} \; \mathcal{L}_\text{int}(\textbf{x})=\int d^{3}\textbf{x} \; e \Bar{\psi}\gamma^{\mu}\psi A_{\mu}.
    \label{Hint}
\end{equation}
We first evaluate the fermion current $\Bar{\psi}\gamma^{\mu}\psi$ in terms of the fermion field's creation and annihilation operators, $\hat{a}$ and $\hat{a}^\dagger$, respectively: 
\begin{equation}
\begin{split}
    \Bar{\psi}\gamma^{\mu}\psi&= \sum_{s,s'}\int \frac{d^3 \textbf{p}_{i} d^3 \textbf{p}_{f}}{(2\pi)^6 \sqrt{4E_{p_{i}}E_{p_{f}}}} \\
    &\times \; \hat{a}_{p_{f}}^{s' \; \dag}\hat{a}_{p_{i}}^{s} e^{-i(\textbf{p}_{i}-\textbf{p}_{f}) \cdot \textbf{x}} \Bar{u}(\boldsymbol{p}_{f})^{s'} \gamma^{\mu} u(\boldsymbol{p}_{i})^{s},    
\end{split}
    \label{ion}
\end{equation}
where 
$u$ is the spinor that the creation/annihilation operators are associated to, $\textbf{p}$ are the momenta of the particle and $s=1,2$ labels the two component spinors. Here, we will assume that $u({\boldsymbol p})^{s}$ corresponds to the fermionic field of the environment. Moreover, we also assume that there is no negative-frequency spinors (anti-particle with associated $b$-creation and annihilation operators) in the on-shell state.

The other quantum field that appears in Eq.~(\ref{EM lagrangian}) is a generic electromagnetic field $A_{\mu}(\textbf{x})$. In our case this term can be related to the crystal in spatial superposition by considering its fermion current $J_{\mu}(\textbf{x})$ associated with the object. In particular, from the Maxwell's equations, we have:
\begin{equation}
    A_{\mu}(\textbf{x})= \Box^{-1} J_{\mu}(\textbf{x}),
    \label{four}
\end{equation}
where $\Box=\eta_{\mu\nu}\partial^{\mu}\partial^{\nu}$ is the D'Alembertian operator and the inverse of the virtual photon propagator. 
Substituting $A_{\mu}$ into Eq.~\eqref{EM lagrangian}, we can study the interaction between two fermions by the exchange of a virtual photon, e.g., between the environmental and crystal fermions. 

We will consider the crystal to be a heavy fermion, and denote the spinor as $U({\boldsymbol k})^{r}$, where $r=1,2$ labels the two component spinors.
In the momentum space Eq.~\eqref{four} becomes:
\begin{equation}
    \Tilde{A}_{\mu}(\textbf{q})=\frac{-i\;g_{\mu\nu}}{q^2} \Tilde{J}^{\nu}(\textbf{q}),
\end{equation}
where $\textbf{q}$ represents the transferred momentum.


%

The fermion current of Eq.~\eqref{four} can be expressed by the crystal's fermion $\Psi$:
\begin{equation}
\begin{split}
    J^{\mu}(\textbf{x})&= \Bar{\Psi} \gamma^{\mu} \Psi
    =\int \frac{d^3 \textbf{k}_{i} d^3 \textbf{k}_{f}}{(2\pi)^6 \sqrt{4E_{k_{i}}E_{k_{f}}}}\hat{c}_{k_{f}}^{r' \; \dag}\hat{c}_{k_{i}}^{r} \\ &\times \sum_{r,r'} e^{-i(\textbf{k}_{i}-\textbf{k}_{f}) \cdot \textbf{x}} \overline{U}(\boldsymbol{k}_{f})^{r'} \gamma^{\mu} U(\boldsymbol{k}_{i})^{r}
\end{split}
\end{equation}
where $\hat{c}$ and $\hat{c}^\dagger$ are respectively the creation and annihilation operators associated to the spinor $U(\boldsymbol{k})^{r}$ of the fermion field that describes the crystal.
This leads to:
\begin{equation}
\begin{split}
    \Tilde{J}^{\nu}(\textbf{q})=\int d^3 \textbf{x}\; e^{i\textbf{q} \cdot \textbf{x}} J^{\nu}(\textbf{x})= \int \frac{d^3 \textbf{k}_{i} d^3 \textbf{k}_{f}}{(2\pi)^6 \sqrt{4E_{k_{i}}E_{k_{f}}}} \\
    \times \sum_{r,r'} \hat{c}_{k_{f}}^{r' \; \dag}\hat{c}_{k_{i}}^{r} \overline{U}(\boldsymbol{k}_f)^{r'} \gamma^{\mu} U(\boldsymbol{k}_{i})^{r}\; \delta^3(\textbf{q}+\textbf{k}_{i}-\textbf{k}_{f}),
\end{split}
\label{FTcurrent}
\end{equation}
where the Dirac's delta $\delta^3(\textbf{q}+\textbf{k}_{i}-\textbf{k}_{f})$ encodes the conservation of momentum during the scattering process.

Plugging Eq.~\eqref{FTcurrent} into Eq.~\eqref{Hint} through:
\begin{equation}
\begin{split}
 A_{\mu}(\textbf{x})&=\int d^3 \textbf{q}\;e^{-i \textbf{q} \cdot \textbf{x}} \Tilde{A}_{\mu}(\textbf{q})\\
 &=\int d^3 \textbf{q}\;e^{-i\textbf{q} \cdot \textbf{x}} \frac{-i\;g_{\mu\nu}}{q^2} \Tilde{J}^{\nu}(\textbf{q}),    
\end{split}
\end{equation}
one obtains:
\begin{align}
     \Hat{H}_\text{int}&=(2\pi)^3 e\int \frac{d^3 \textbf{p}_{i} d^3 \textbf{p}_{f} d^3 \textbf{k}_{i} d^3 \textbf{k}_{f}}{(2\pi)^{12} \sqrt{16\;E_{p_{i}} E_{p_{f}} E_{k_{i}} E_{k_{f}}}} \nonumber\\ 
     &\times \sum_{\substack{s,s' \\ r,r'}}
    \hat{a}_{p_{f}}^{s' \; \dag}\hat{a}_{p_{i}}^{s} \; \hat{c}_{k_{f}}^{r' \; \dag}\hat{c}_{k_{i}}^{r} \; \delta^3(\textbf{k}_{i}+\textbf{p}_{i}-\textbf{k}_{f}-\textbf{p}_{f}) \nonumber\\ 
    &\times \;
    \Bar{u}(\boldsymbol{p}_{f})^{s'} \gamma^{\mu} u(\boldsymbol{p}_{i})^{s} \frac{-i\;g_{\mu\nu}}{q^2} \overline{U}(\boldsymbol{k}_{f})^{r'} \gamma^{\nu} U(\boldsymbol{k}_{i})^{r},
\end{align}
where now the transferred momentum 
$\textbf{q}$ appearing in the propagator is explicitly given by $\textbf{q}=\textbf{k}_{f}-\textbf{k}_{i}$ because of the Dirac's delta appearing in Eq.~\eqref{FTcurrent}.

Notice that the last term in the RHS is exactly the quantum electrodynamics (QED) matrix element for the fermion-fermion interaction $\mathcal{M}_{p_{i},p_{f},k_{i},k_{f}}^{s,s',r,r'}$~\cite{Schwartz:2014sze}:
\begin{equation}
    \begin{split}
     &\Hat{H}_\text{int}=(2\pi)^3 e\int \frac{d^3 \textbf{p}_{i} d^3 \textbf{p}_{f} d^3 \textbf{k}_{i} d^3 \textbf{k}_{f}}{(2\pi)^{12} \sqrt{16\;E_{p_{i}} E_{p_{f}} E_{k_{i}} E_{k_{f}}}} \sum_{\substack{s,s',r,r'}} \\&
     \hat{a}_{p_{f}}^{s' \; \dag}\hat{a}_{p_{i}}^{s} \; \hat{c}_{k_{f}}^{r' \; \dag}\hat{c}_{k_{i}}^{r} \delta^3(\textbf{k}_{i}+\textbf{p}_{i}-\textbf{k}_{f}-\textbf{p}_{f}) \; 
    \mathcal{M}_{p_{i},p_{f},k_{i},k_{f}}^{s,s',r,r'}.
\end{split}
\label{28}
\end{equation}
with~\footnote{Further note that the Dirac delta $\delta^3(\textbf{k}_{i}+\textbf{p}_{i}-\textbf{k}_{f}-\textbf{p}_{f})$ appearing in Eq.~\eqref{28} encodes the conservation of momentum between the initial and the final states of the scattering diagram between the environmental fermion and the crystal's fermion, i.e. $\textbf{k}_{i}+\textbf{p}_{i}=\textbf{k}_{f}+\textbf{p}_{f}$.}
\begin{equation}\label{eq:scat_matrix_el}
\begin{split}
  \mathcal{M}_{p_{i},p_{f},k_{i},k_{f}}^{s,s',r,r'} =& \Bar{u}(\boldsymbol{p}_{f})^{s'} \gamma^{\mu} u((\boldsymbol{p}_{i}))^{s} \\&
  \times \frac{-i\;g_{\mu\nu}}{q^2} \overline{U}(\boldsymbol{k}_{f})^{r'} \gamma^{\nu} U(\boldsymbol{k}_{i})^{r}  .  
\end{split} 
\end{equation}
%

\subsection{Hamiltonian construction}\label{subsec:hamiltonian}

At this point, one can express the creation and annihilation operators ($\hat{c}$'s and $\hat{a}$'s) in terms of one particle states ($\ket{\textbf{k},r}$ and $\ket{\textbf{p},s}$ respectively~\footnote{Here, the notation $\ket{\textbf{k},r}$ means formally $ \ket{\textbf{k}} \otimes \ket{r} \equiv \ket{\textbf{k},r}$, i.e. the state can be separated in the momentum space and the spin space.}) through the relations:
\begin{equation}
\ket{\textbf{k},r}=\sqrt{2 E_{k}}\hat{c}_{k}^{r\; \dag} \ket{0}_{S}\,;~~~
\ket{\textbf{p},s}=\sqrt{2 E_{p}}\hat{a}_{p}^{s \; \dag} \ket{0}_{\mathcal{E}} 
\end{equation}
where $\ket{0}_{\mathcal{E}}$ and $\ket{0}_{S}$ represent respectively the vacuum of the environmental particle and the vacuum of the crystal. In particular, the operator $\hat{A} \equiv \hat{a}_{p_{f}}^{s' \; \dag}\hat{a}_{p_{i}}^{s}$ \textit{destroys} an environmental particle with initial momentum $\boldsymbol{p}_{i}$ and initial spin component $s$ and \textit{creates} an environmental particle with final momentum $\boldsymbol{p}_{f}$ and final spin component $s'$. This means that the action of this operator can be written as:

\begin{equation}
    \hat{A} : \ket{\textbf{p}_{i},s} \; \longrightarrow \ket{\textbf{p}_{f},s'}.
    \label{operator_action}
\end{equation}

Now, we can write the creation operator $\hat{a}_{p}^{s \; \dagger}$ (which acts on the vacuum as $\ket{\textbf{p},s}=\sqrt{2 E_{p}}\hat{a}_{p}^{s \; \dag} \ket{0}_{\mathcal{E}} $) in the momentum basis representation $\left\{ \ket{\textbf{p},s} \right\}$ as:
\begin{equation}
    \hat{a}_{p}^{s \;\dagger}= \frac{1}{\sqrt{2 E_{p}}} \ket{\textbf{p},s} \bra{0}_{\mathcal{E}}.
    \footnote{One can indeed verify that the action on the vacuum $\ket{0}_{\mathcal{E}} $ of this operator is $\frac{1}{\sqrt{2 E_{p}}} \ket{\textbf{p},s}_{\;\mathcal{E}} \braket{0}_{\mathcal{E}} = \frac{1}{\sqrt{2 E_{p}}} \ket{\textbf{p},s} \cdot 1= \frac{1}{\sqrt{2 E_{p}}} \ket{\textbf{p},s}$, which is exactly the definition of creation operator.}
\end{equation}
Because of this property, the operator $\hat{A}$ becomes:
\begin{equation}
\begin{split}
    \hat{A} \equiv \hat{a}_{p_{f}}^{s' \; \dag}\hat{a}_{p_{i}}^{s} = \frac{1}{\sqrt{4 E_{p_{i}} E_{p_{f}}}} \ket{\textbf{p}_{f},s'} \bra{\textbf{p}_{i},s},    
\end{split}
\end{equation}
which satisfies Eq.(\ref{operator_action}). The same relations hold for the crystal field, with $\ket{\textbf{p},s} \rightarrow \ket{\textbf{k},r}$, $\hat{a} \rightarrow \hat{c}$ and $\ket{0}_{\mathcal{E}} \rightarrow \ket{0}_{S} $. This means that Eq.(\ref{28}) gives the following form for the Hamiltonian interaction $\hat{H}_\text{int}$:
%
\begin{equation}
    \begin{split}
     \Hat{H}_\text{int}&=\int \frac{d^3 \textbf{p}_{i} d^3 \textbf{p}_{f} d^3 \textbf{k}_{i} }{(2\pi)^{9} 16\;E_{p_{i}} E_{p_{f}} E_{k_{i}} E_{k_{i}+p_{i}-p_{f}}} \\
     &\times \sum_{\substack{s,s',r,r'}}\mathcal{M}_{p_{i},p_{f},k_{i},k_{i}+p_{i}-p_{f}}^{s,s',r,r'} \; \ket{\textbf{p}_{f},s'} \bra{\textbf{p}_{i},s} \\ 
     &\otimes e^{i(\textbf{p}_{i}-\textbf{p}_{f}) 
\cdot \hat{\textbf{x}}} \ket{\textbf{k}_{i},r'} \bra{\textbf{k}_{i},r},
\end{split}
\label{followHint}
\end{equation}
where $\delta^3(\textbf{k}_{i}+\textbf{p}_{i}-\textbf{k}_{f}-\textbf{p}_{f})$ has been absorbed by performing the integral over $d^{3} \textbf{k}_{f}$. The vector state $\ket{\textbf{k}_{f}} \rightarrow \ket{\textbf{k}_{i}+\textbf{p}_{i}-\textbf{p}_{f}}$ has been expressed in terms of the translation operator, which generates the translations in the momenta space through the position operator $\hat{\textbf{x}}$, i.e. $\ket{\textbf{k}_{i}+\textbf{p}_{i}-\textbf{p}_{f}}=e^{i(\textbf{p}_{i}-\textbf{p}_{f}) \cdot \hat{\textbf{x}}} \ket{\textbf{k}_{i}}$.

To simplify the expression of the interaction Hamiltonian in Eq.~\eqref{followHint} we make some assumptions:
\begin{itemize}
\item The environmental particle's momenta are assumed to be very small, due to a low ambient temperature in the original experiment, we will quantify that later.

\item  The crystal mass is assumed to be much heavier than the environmental particle's mass, so that its momentum does not change during the scattering process with the environmental particles.

\end{itemize}

This means that $\abs{\mathbf{p}_{i}},\abs{\mathbf{p}_{f}} \ll M$, where $M$ is the crystal's mass. As a result $\abs{\mathbf{k}_{i}}, \abs{\mathbf{k}_{f}} \sim 0$.
These assumptions are valid for most types of experiments, so the application of their Hamiltonian is kept general while the approximations simplify the expression.
These assumptions allow us to approximate the matrix element in Eq.\eqref{followHint} as 
\begin{equation}
\mathcal{M}_{p_{i},p_{f},k_{i},k_{i}+p_{i}-p_{f}}^{s,s',r,r'} \sim \mathcal{M}_{p_{i},p_{f},0,0}^{s,s',r,r'}\,.
\end{equation}

Since the crystal is heavy with respect to the ambient energy, we can further approximate $2 E_{k_{i}}2 E_{k_{i}+p_{i}-p_{f}} \sim 4ME_{k_{i}}$.
Finally, integrating over $\boldsymbol{k}_{i}$ and using the identity $\int \frac{d^3 \textbf{k}_{i}}{(2\pi)^3 2E_{k_{i}}} \ket{\textbf{k}_{i}}\bra{\textbf{k}_{i}}=\mathbb{I}$, we can obtain the final expression for Eq.\eqref{Hint}:
\begin{equation}
\begin{aligned}
    \Hat{H}_{int}&=\int \frac{d^3 \textbf{p}_{i} d^3 \textbf{p}_{f}}{(2\pi)^6 2E_{p_{i}}2E_{p_{f}}} \sum_{\substack{s,s',r,r'}} \frac{\mathcal{M}_{p_{i},p_{f}}^{s,s',r,r'}}{2M} \\
    &\times \ket{\textbf{p}_{f},s'} \bra{\textbf{p}_{i},s} \otimes e^{i(\textbf{p}_{i}-\textbf{p}_{f}) 
    \cdot \hat{\textbf{x}}} \, \ket{r'} \bra{r}.    
\end{aligned}
    \label{finalHint}
\end{equation}
%


\subsection{Born-Markov Master Equation}\label{subsec:master_eqn}

To find a Master Equation that described the decoherence of the crystal due to the environment, we start by supposing that the interaction Hamiltonian has the general form \cite{Schlosshauer:2014pgr, joos2003decoherence}:
\begin{equation}
    \hat{H}_\text{int}=\sum_{\alpha} \hat{S}_{\alpha} \otimes \hat{E}_{\alpha},
   \label{ham}
\end{equation}
where $\hat{S}_{\alpha}$ and $\hat{E}_{\alpha}$ represent all the degrees of freedom of the matter-wave interferometer and the environment respectively and $\alpha$ represents schematically the sums over the momenta $\textbf{p}$ and the two component spinors $s$~\footnote{ More explicitly, from Eq.\eqref{finalHint} we have  $\sum_{\alpha} \equiv \int \frac{d^3 \textbf{p}_{i} d^3 \textbf{p}_{f}}{(2\pi)^6 2E_{p_{i}}2E_{p_{f}}} \sum_{\substack{s,s',r,r'}}$}.
From Eq.\eqref{finalHint}, one can find the explicit expression for $\hat{S}_{\alpha}$ and $\hat{E}_{\alpha}$:
\begin{align}
        \hat{S}_{\alpha}&=\frac{e^{i(\textbf{p}_{i}-\textbf{p}_{f}) 
        \cdot \hat{\textbf{x}}}}{2M} \ket{r'} \bra{r}\nonumber  \\
        \hat{E}_{\alpha}&=\mathcal{M}_{p_{i},p_{f}}^{s,s',r,r'} \ket{\textbf{p}_{f},s'} \bra{\textbf{p}_{i},s}
    \label{SandE}
\end{align}
which matches the result in Ref.~\cite{Kilian:2022kgm} which described the scattering between a neutrino and a heavy nucleus, mediated by the weak interaction. 
The difference is that in Ref.~\cite{Kilian:2022kgm} the matrix element $\mathcal{M}_{p_{i},p_{f}}^{s,s',r,r'}$ is defined by the scattering via the weak interaction, while we have defined it in terms of the electromagnetic interaction as in Eq.~\eqref{eq:scat_matrix_el}.

%
The master equation in the Born-Markov approximation is given by~\cite{Schlosshauer:2019ewh}:
\begin{equation}
    \begin{aligned} \frac{d \rho_S}{d t} & =-\frac{i}{\hbar}\left[H_S, \rho_S\right]-\bigg\{\int_0^{\infty} d \tau \sum_{\alpha \beta} C_{\alpha \beta}(-\tau) \\ & \times\left[S_\alpha S_\beta(-\tau) \rho_S-S_\beta(-\tau) \rho_S S_\alpha\right]+\text { H.c. }\bigg\} \end{aligned},
    \label{master}
\end{equation}
where
\begin{equation}
    C_{\alpha \beta}=\frac{1}{\hbar^2} \operatorname{Tr}\left[\rho_E E_\alpha E_\beta(-\tau)\right]
    \label{Cab}
\end{equation}
and
\begin{equation}
    \begin{aligned} S_\beta(-\tau) & =e^{-\frac{i H_S \tau}{\hbar}} S_\beta e^{\frac{i H_S \tau}{h}}, \\ E_\beta(-\tau) & =e^{-\frac{i H_E \tau}{\hbar}} E_\beta e^{\frac{i H_E \tau}{h}} . \end{aligned}
\end{equation}
%
In Eq.~\eqref{master} we have made the Born and Markov approximations.
The Born approximation considers the environment to be much larger than the system, and the coupling between $S$ and $\mathcal{E}$ to be weak enough that it is possible at all times to write the composite $S+\mathcal{E}$ system as a tensor product:
    \begin{equation}
        \hat{\rho}_{S+\mathcal{E}}(t) \approx \hat{\rho}_{S}(t) \otimes \hat{\rho}_{\mathcal{E}},
    \end{equation}
where $\hat{\rho}_{\mathcal{E}}$ is approximately constant at all times.

The Markov approximation supposes that the memory effects in the environment are negligible, i.e. any effect that the system has on the environment decays rapidly compared to the evolution of the environment $\mathcal{E}$ itself.

Additionally we assume that the time evolution of the operator $S_{\beta}$ can be neglected with respect to the correlation time scale of the environment, which is applicable for example when the centre-of-mass of the crystal is trapped in a very low-frequency trap, as we will see in th examples of the QGEM experiment, see~\cite{Marshman:2023nkh}.

We also assume that the \textit{unitary} time evolution of the crystal (given by $-{i}\hbar^{-1}\left[H_S, \rho_S\right]$ of Eq.~\eqref{master}) is much slower than the \textit{non-unitary} time evolution (given by the second term of the right-hand-side of Eq.~\eqref{master}), which corresponds to changes of the system entirely due to the decoherence. 

This means that the decoherence due to the presence of the environment modifies the state of the system faster than any \textit{free} evolution of the system itself. For this reason, in the derivation of the decoherence rate below, we will neglect the unitary term $-{i}{\hbar^{-1}}\left[H_S, \rho_S\right]$.

We consider the environmental particles to have a (normalized) wave function with a localized momentum $\mathbf{p}_{0}$. We represent the environmental particle using a Gaussian wavepacket centered in $\mathbf{p}_{0}$ in the momentum space and with a width $\Tilde{\sigma}$ and with spin component $s_{0}$:
\begin{equation}
    \begin{aligned} |\psi_E\rangle & =\int \frac{d^3 \mathbf{p}}{(2 \pi)^3 \sqrt{2 E_p}} \Tilde{\psi}(p)|\mathbf{p},s_{0}\rangle \\ & =\int \frac{d^3 p}{(2 \pi)^3 \sqrt{2 E_p}} \frac{(2 \pi)^{3 / 2}}{\left( \pi \Tilde{\sigma}^2\right)^{3 / 4}} e^{-\frac{|\mathbf{p}-\mathbf{p}_0|^2}{2 \Tilde{\sigma}^2}}|\mathbf{p},s_{0}\rangle . \end{aligned}
    \label{wfp}
\end{equation}
From this wavefunction for the environmental particle we compute it's density matrix $\hat{\rho}_{E}=\ket{\Psi_E}\bra{\Psi_E}$.

Now, we can substitute this density matrix in Eq.~\eqref{Cab} in order to compute the coefficients $C_{\alpha,\beta}$. Notice that Eq.~\eqref{Cab} depends on two indices, $\alpha$ and $\beta$, so it is important to find a suitable notation for both of them. In particular, we will define $ \left\{ \alpha \right\} \equiv \left\{ \textbf{p}_{i},\textbf{p}_{f}, s,s',r,r' \right\} $ and $\left\{ \beta \right\} \equiv \left\{ \textbf{p}_i^{\prime},\textbf{p}_f^{\prime}, n,n',m,m' \right\}$. At this point, we are ready to express the coefficients $C_{\alpha,\beta}$ explicitly:
\begin{widetext}
\begin{align}
    C_{\alpha,\beta}&=\Tr[\rho_{E}E_{\alpha}E_{\beta}(-\tau)]=\left\langle E_{\alpha}E_{\beta}(-\tau) \right\rangle_{\psi_{E}} \\
    &=(2 \pi)^3 \mathcal{M}_{p_i, p_f}^{s,s',r,r'} \left(\mathcal{M}_{p_f^{\prime}, p_i^{\prime}}^{n,n',m,m'}\right)^{\ast} 2 E_{p_i} \delta^3\left(\textbf{p}_i-\textbf{p}_f^{\prime}\right) \delta_{s n'}
    \bra{\psi_{E}}e^{-i\left(E_{p_i^{\prime}}-E_{p_f^{\prime}}^{\prime}\right) \tau}\ket{\textbf{p}_{f}}\left\langle \textbf{p}_i^{\prime} \mid \psi_{E}\right\rangle \\
    &=(2 \pi)^9 \pi^{-3/2} \tilde{\sigma}^3 \mathcal{M}_{p_i, p_f}^{s,s',r,r'} \left(\mathcal{M}_{p_f^{\prime}, p_i^{\prime}}^{n,n',m,m'}\right)^{\ast} 2 E_{p_i} 
    \delta^3\big(\textbf{p}_i-\textbf{p}_f^{\prime}\big)\delta_{s n'} e^{-i\left(E_{p_i^{\prime}}-E_{p_f^{\prime}}\right) \tau} 2 E_{p_i^{\prime}} \nonumber\\ &\qq{}\times\delta_{s_{0} s'} \delta_{n s_{0}} \delta^3\left(\textbf{p}_i^{\prime}-\textbf{p}_0\right)\delta^3\left(\textbf{p}_0-\textbf{p}_f\right),\label{Cabexplic}
\end{align}
\end{widetext}
where factors like $\delta_{s_{0} s'}$ indicate Kronecker's deltas, because the spin components have discrete values~\footnote{The two component spinors obey the Kronecker delta, e.g. $\xi^{s_0 \dagger}\xi^{s'}=\delta^{s_0s'}\equiv \delta_{s_0s'}$, where $\xi$ is the two component spinor.}. Moreover, in Eq.~\eqref{Cabexplic} we have used the following two properties:
\begin{equation}\label{eq:props}
    \begin{cases}
        \bra{\textbf{p},s}\ket{\textbf{k},r}=(2\pi)^3 (2E_{p}) \delta^3(\textbf{p}-\textbf{k}) \delta_{s r} \\
        \delta^3(\textbf{p})=\lim_{\Tilde{\sigma} \to 0} \frac{1}{\Tilde{\sigma}^3\pi^{3/2}} e^{-|\textbf{p}|^2/\Tilde{\sigma}^2}
    \end{cases}.
\end{equation}
%
The second property in Eq.~\eqref{eq:props} represents the assumption that the Gaussian wavepacket of the environmental particle is sharp enough that can be considered as a Dirac's delta, i.e. the particle has a well-defined momentum $\textbf{p}_{0}$.
 
We are now ready to find an explicit expression for the temporal evolution of the density matrix $\hat{\rho}_{S} $ as given in Eq.~\eqref{master}. In particular, plugging Eq.~\eqref{SandE} into the second term of the right-hand-side of Eq.~\eqref{Cab} (and neglecting the unitary term $-\frac{i}{\hbar}\left[H_S, \rho_S\right]$ as discussed above), one obtains:
\begin{widetext}
\begin{equation}
\begin{split}
    \frac{d \hat{\rho}_{S}}{dt}=&-\frac{\pi^{-3/2} \Tilde{\sigma}^3}{4M^2} \int d\tau \frac{d^3 \textbf{p}_{i}d^3 \textbf{p}_{f}d^3 \textbf{p}'_{i}d^3 \textbf{p}'_{f}}{(2\pi)^3 16 E_{p_{i}} E_{p_{f}}E_{p'_{i}}E_{p'_{f}}} 4 E_{p_{i}} E_{p'_{i}} \sum_{\substack{s,s',r,r'\\
    n,n',m,m'}}\mathcal{M}_{p_{i},p_{f}}^{s,s',r,r'} \Big(\mathcal{M}_{p'_{i},p'_{f}}^{n,n',m,m'}\Big)^{*} \delta^3(\textbf{p}_{i}-\textbf{p}'_{f}) \\
    & \times \delta_{s n'} e^{-i(E_{p'_{i}}-E_{p'_{f}})\tau}\delta^3(\textbf{p}_{0}-\textbf{p}_{f}) \delta^3(\textbf{p}'_{i}-\textbf{p}_{0}) \delta_{s_{0} s'} \delta_{n s_{0}} \big\{-e^{i(\textbf{p}'_{i}-\textbf{p}'_{f})\cdot \hat{\textbf{x}}} \ket{m'} \bra{m} \hat{\rho}_{S} e^{i(\textbf{p}_{i}-\textbf{p}_{f}) \cdot \hat{\textbf{x}}}\ket{r'} \bra{r}
    \\& \;\;\;\;\;\;\;\;\;\;\;\;\;\;\;\;\;\;\;\;\;\;\;\;\;\;\;\;\;\;\;\;\;\;\;\;\;\;\;\;\;\;\;\;\;\;\;\;\;\;\;\;\;\;\;\;\;\;\;\;\;\;\;\;\;\;\;\;\;\;\;\;\;\;\;\;\;\;\;\;\;\;\;+e^{i(\textbf{p}_{i}-\textbf{p}_{f})\cdot \hat{\textbf{x}}} \ket{r'} \bra{r} e^{i(\textbf{p}'_{i}-\textbf{p}'_{f}) \cdot \hat{\textbf{x}}} \ket{m'} \bra{m}\hat{\rho}_{S}+\text{H.c.} \big\},
\end{split} \label{longequation}
\end{equation}    
\end{widetext}
which, after solving the Dirac and Kronecker deltas, leads to:
\begin{widetext}
\begin{align} \label{twentysix}
    \frac{d \hat{\rho}_{S}}{dt}=&-\frac{\pi^{-3/2} \Tilde{\sigma}^3}{64 \pi^2 M^2 E_{p_{0}}} \int \frac{d^3\textbf{p}'_{f}}{E_{p'_{f}}} \times \sum_{\substack{s,r,r'\\m,m'}}\mathcal{M}_{p_{0},p'_{f}}^{s,s_{0},r,r'}\Big(\mathcal{M}_{p_{0},p'_{f}}^{s_{0},s,m,m'}\Big)^{*}\delta(E_{p_{0}}-E_{p'_{f}})  \nonumber\\ 
    &\times  \big \{-\ket{m'}\bra{m}e^{i(\textbf{p}_{0}-\textbf{p}'_{f})\cdot \hat{\textbf{x}}} \hat{\rho}_{S} e^{i(\textbf{p}'_{f}-\textbf{p}_{0}) \cdot \hat{\textbf{x}}}\ket{r'}\bra{r}
    +\delta_{r m'}\ket{r'} \bra{m} e^{i(\textbf{p}'_{f}-\textbf{p}_{0}) \cdot \hat{\textbf{x}}}  e^{i(\textbf{p}'_{i}-\textbf{p}'_{f}) \cdot \hat{\textbf{x}}} \hat{\rho}_{S}+ \text{H.c.} \big\},     
\end{align}
\end{widetext}
where the one-dimensional Dirac's delta $\delta(E_{p_{0}}-E_{p'_{f}})$ comes from the fact that we have integrated over 
$d \tau$~\footnote{ $\delta(E_{p'_{i}}-E_{p'_{f}})=\int d \tau \; e^{-i(E_{p'_{i}}-E_{p'_{f}})\tau}$}, while simultaneously solving the other Dirac delta $\delta^3(\textbf{p}'_{i}-\textbf{p}_{0})$ appearing in Eq.\eqref{longequation}, therefore giving $\delta(E_{p'_{i}}-E_{p'_{f}})  \rightarrow \delta(E_{p_{0}}-E_{p'_{f}})$.
At this point, solving $\delta(E_{p_{0}}-E_{p'_{f}})$ and using $dp'_{f}=({E_{p'_{f}}}/{p'_{f}})dE_{p'_{f}}$ (we denote $\abs{\textbf{p}'_{f}}$ as $p'_{f}$), gives:
\begin{widetext}
\begin{equation}
\begin{split}
     \frac{d \hat{\rho}_{S}}{dt}=&-\frac{\pi^{-3/2} \Tilde{\sigma}^3}{64 \pi^2 M^2 } \frac{p_{0}}{E_{p_{0}}}  \int d\Omega '\sum_{\substack{s,r,r'\\m,m'}}\mathcal{M}_{p_{0},p'_{f}}^{s,s_{0},r,r'}\Big(\mathcal{M}_{p_{0},p'_{f}}^{s_{0},s,m,m'}\Big)^{*} \\&
     \times \big\{ -\ket{m'} \bra{m}e^{i(\textbf{p}_{0}-\textbf{p}'_{f})\cdot\hat{\textbf{x}}} \hat{\rho}_{S} e^{i(\textbf{p}'_{f}-\textbf{p}_{0})\cdot\hat{\textbf{x}}} 
     \otimes \ket{r'} \bra{r}
    + \delta_{r m'} \ket{r'} \bra{m}\hat{\rho}_{S}+H.c. \big\},
\end{split}
\label{eq:master2}
\end{equation}
\end{widetext}
where $d \Omega '$ indicates the orientation angle of the final momentum vector $d^{3} \textbf{p}_{f}= d \Omega ' dp_{f} p_{f}^2$.
Note that $E_{p_{0}} = E_{p'_{f}}$ gives $|\textbf{p}_{0}|=|\textbf{p}'_{f}|$.
This means also that the matrix element $\mathcal{M}_{p_{0},p'_{f}}$ will now depend only on $|\textbf{p}_{0}|$ and the angle between $\textbf{p}_{0}$ and $\textbf{p}'_{f}$, i.e. on $\Omega '$.
We can also rewrite the ${p_{0}}/{E_{p_{0}}}$ term as ${p_{0}}/{E_{p_{0}}}=v_{0}$~\footnote{This relation can be easily derived from well known relativistic expressions, i.e. $\frac{p_{0}}{E_{p_{0}}}=\frac{\frac{mv_{0}}{\sqrt{1-v_{0}^2}}}{\sqrt{p_{0}^2+m^2}}=\frac{\frac{mv_{0}}{\sqrt{1-v_{0}^2}}}{\sqrt{\frac{m^2v_{0}^2}{1-v_{0}^2}+m^2}}=v_{0}$.}.

The factor $\textbf{p}'_{f}-\textbf{p}_{0}$ at the exponent of the right-hand side of Eq.~\eqref{eq:master2} can be rewritten as $|\textbf{p}_{0}|(\hat{n}'-\hat{n}_{0})$, where $\hat{n}'$  and $\hat{n}_{0}$ are the final and initial direction of the scattered environmental particle (we will express their associated angles with, respectively, $\Omega '$ and $\Omega_{0}$). 
We can now use this notation to rewrite the matrix $\mathcal{M}_{p_{0},p'_{f}}^{s,s_{0},r,r'}$ in terms of the scattering angles $\Omega '$ and $\Omega_{0}$: in fact, it is always possible to find a suitable parameterization
of the momenta $\textbf{p}_{0}$ and $\textbf{p}'_{f}$ in terms of the COM energy and their angles $\Omega '$ and $\Omega_{0}$. In particular, from now on we will write $\mathcal{M}_{p_{0},p'_{f}}^{s,s_{0},r,r'} \rightarrow \mathcal{M}_{\Omega ',\Omega_{0},E_{0}}^{s,s_{0},r,r'}$~\cite{Kilian:2022kgm}. In this way, it is possible to have the matrix $\mathcal{M}$ with an explicit dependence on the integration variable appearing in Eq.\eqref{eq:master2}, i.e. $\Omega '$.

We can thus rewrite the time evolution equation for $\hat{\rho}_{S}$ as:
\begin{equation}
    \begin{split}
     \frac{d \hat{\rho}_{S}}{dt}=&-\frac{\pi^{-3/2} \Tilde{\sigma}^3}{64 \pi^2 M^2 } v_{0}\\&
     \times \int d\Omega '\sum_{\substack{s,r,r'\\m,m'}}\mathcal{M}_{\Omega ',\Omega_{0},E_{0}}^{s,s_{0},r,r'}\Big(\mathcal{M}_{\Omega ',\Omega_{0},E_{0}}^{s_{0},s,m,m'}\Big)^{*} \\&
     \times \big\{ -\ket{m'} \bra{m}e^{i(\textbf{p}_{0}-\textbf{p}'_{f})\cdot\hat{\textbf{x}}} \hat{\rho}_{S} e^{i(\textbf{p}'_{f}-\textbf{p}_{0})\cdot\hat{\textbf{x}}} \\&
     \otimes \ket{r'} \bra{r}
    + \delta_{r m'} \ket{r'} \bra{m}\hat{\rho}_{S}+H.c. \big\}. 
\end{split}
\label{42}
\end{equation}
The factor $ \pi^{-3/2} \Tilde{\sigma}^3 $ is proportional to the inverse of the volume ${1}/{V}$, see Appendix~\ref{App}. Using this proportionality in Eq.\eqref{42} a term like ${v_{0}}/{V}=F_{1}$ will appear, which is the flux associated with one environmental particle.

Now, to obtain the matrix elements of $S$, we need to perform the bracket of the operator $\hat{\rho}_{S}$ with \textit{both} the spatial degrees of freedom $\big\{ \ket{\textbf{x}} \big\}$ and the spin ones $\big\{ \ket{r} \big\}$. Therefore, contracting $\hat{\rho}_{S}$ with two generic spin vectors and calling them $\ket{r_{i}}$ and $\ket{r_{f}}$, the RHS of Eq.~(\ref{42}) becomes:
\begin{equation}
    \begin{split}
    \frac{d \hat{\rho}_{S}}{dt}=&- \frac{(2\pi)^{3/2}F_{1} }{64 \pi^2 M^2 }\\& 
      \times \int d\Omega '\sum_{\substack{s,r,r'\\m,m'}}\mathcal{M}_{\Omega ',\Omega_{0},E_{0}}^{s,s_{0},r,r'}\Big(\mathcal{M}_{\Omega ',\Omega_{0},E_{0}}^{s_{0},s,m,m'}\Big)^{*} \\&
      \times \big\{ -\delta_{r_{f} m'} \delta_{r r_{i}} \bra{m}e^{i(\textbf{p}_{0}-\textbf{p}'_{f})\cdot\hat{\textbf{x}}} \hat{\rho}_{S} e^{i(\textbf{p}'_{f}-\textbf{p}_{0})\cdot\hat{\textbf{x}}} \ket{r'}\\& + \delta_{r m'} \delta_{r_{f} r'}\bra{m}\hat{\rho}_{S}\ket{r_{i}}+H.c. \big\}. 
    \end{split} \label{32}
\end{equation}
Note from Eq.\eqref{32}, the terms inside the final curly brackets has four Kronecker deltas, which will annihilate some indices in the summation~\footnote{ In particular, the first two Kronecker deltas $\delta_{r_{f} m'} \delta_{r r_{i}}$ will give $\sum_{\substack{s,r,r'\\m,m'}} \rightarrow \sum_{\substack{s,r',m}}$, while the second two kronecker deltas $\delta_{r m'} \delta_{r_{f} r'}$ will give $\sum_{\substack{s,r,r'\\m,m'}} \rightarrow \sum_{\substack{s,r,m}}$.}. We can rename the $r'$ index in the first summation $\sum_{\substack{s,r',m}} \rightarrow \sum_{\substack{s,r,m}}$ to obtain the same indices of the second one, giving us with the unique final summation $\sum_{\substack{s,r,m}}$.
%
%
We can now make some approximation to simplify Eq.\eqref{32}. In particular, we will assume that the interaction happens in the \textit{non-relativistic} regime. This is a fair assumption, if one considers the scenario in the QGEM proposal \cite{Bose:2017nin}, where the ambient temperature is very low to maintain coherence of the interferometer, i.e. $T \sim 1\; \text{K}$, see the section below for the parameters of the QGEM experiment. Therefore, we can use the fundamental properties of the QED at low energy regime, where the spin components are preserved during the scattering process~\footnote{In the non-relativistic limit we have that $p^{\mu} \rightarrow (m,\textbf{p}) $ and thus a generic spinor $u(\textbf{p})= 
\left(\begin{matrix}
   \sqrt{\textbf{p} \cdot \sigma} \xi\\
   \sqrt{\textbf{p} \cdot \Bar{\sigma}} \xi
\end{matrix}\right) \rightarrow \sqrt{m}\left(\begin{matrix}
    \xi\\
    \xi
\end{matrix}\right)$, where $\sigma$ represents the Pauli matrices and $\xi$ is a generic $2$ component spinors. This leads to the following spinor contractions $\sum_{s,s'}\Bar{u}^{s'} u^{s} \rightarrow 2 m \delta_{s s'}$ which is responsible to the existence of the spin conservation term $\delta_{s s'}$ inside the matrix $\mathcal{M}^{s s'}$. For further details, see \cite{tong}.}. This property leads to the following form for the scattering matrix $\mathcal{M}_{s,s_{0}}^{r ,r'} \rightarrow \mathcal{M}\; \delta^{r r'} \delta^{ss_{0}}$. Eq.\eqref{32} thus becomes:
\begin{equation}
    \begin{split}
        \frac{d \hat{\rho}_{S}}{dt}=&- \frac{(2\pi)^{3/2}F_{1} }{64 \pi^2 M^2 } \int d\Omega ' |\mathcal{M}(\Omega ',\Omega_{0},E_{0})|^{2}\\&
        \times \Big\{ -\bra{r_{f}}e^{i(\textbf{p}_{0}-\textbf{p}'_{f})\cdot\hat{\textbf{x}}} \hat{\rho}_{S} e^{i(\textbf{p}'_{f}-\textbf{p}_{0})\cdot\hat{\textbf{x}}} \ket{r_{i}} \\&
        + \bra{r_{f}}\hat{\rho}_{S}\ket{r_{i}}+H.c. \Big\}
    \end{split}
\end{equation}
Now, contracting over the spatial components, $\ket{\textbf{x}}$, and defining $ \bra{\textbf{x},r_{f}} \hat{\rho}_{S} \ket{\textbf{y},r_{i}}
\equiv \rho_{S}^{r_{i},r_{f}}(\textbf{x}, \textbf{y},t)$, we obtain:
\begin{equation}
    \begin{aligned} 
    \frac{d \rho_{S}^{r_{i},r_{f}}}{dt}&(\textbf{x}, \textbf{y},t) =- \frac{2(2\pi)^{3/2}F_{1}}{64 \pi^2 M^2 } \int d\Omega '|\mathcal{M}(\Omega ',\Omega_{0},E_{0})|^2\\&
    \times  \left\{-e^{-ip_{0}\left(\hat{n}'-\hat{n}_{0}\right)\cdot(\textbf{x}-\textbf{y})}+1\right\}\rho_{S}^{r_{i},r_{f}} (\textbf{x}, \textbf{y},t),
    \end{aligned} \label{onepart}
\end{equation}
where we have defined $\rho_{S}^{r_{i},r_{f}}(\textbf{x},\textbf{y},t) \equiv \bra{\textbf{x},r_{f}} \hat{\rho}_{S} \ket{\textbf{y},r_{i}}$. The term $(64 \pi^2 M^2)^{-1}|\mathcal{M}(\Omega ',\Omega_{0},E_{0})|^2$ is exactly the differential cross-section, ${d \sigma}/{d \Omega '}={d \sigma}(\hat{n}_{0},\hat{n}^{'})/{d \Omega '}$, (see Appendix \eqref{Appendix}).
Note that Eq.~\eqref{onepart} has the same form as the result found in Ref.~\cite{Kilian:2022kgm}, but with a different scattering amplitude, there it was the weak interaction, and here it is the electromagnetic interaction.

Solution of the Eq.~\eqref{onepart} will dictate the decoherence of the matter-wave interferometer due to the interaction between one environmental particle and the micro-crystal. It is easy to generalize it to the case when we have $N$ environmental particles.
The flux is then given by: 
\begin{equation}
F=n\Bar{v}=\frac{N}{V}\Bar{v},
\end{equation}
where $\Bar{v}$ is the average velocity of each environmental particle. 
In general, if each particle gives a contribution of the type Eq.~\eqref{onepart} to the decoherence rate, the total contribution will be given by the sum over all possible momenta $\textbf{p}_{0}$ (and velocities $\textbf{v}_{0}$) weighted by a distribution of particles $\mu(\textbf{p}_{0})$ in the momentum space, i.e. a probabilistic distribution such that $\int d^3 \textbf{p}_{0} \; \mu(\textbf{p}_{0})=1$. Assuming that the environment is composed of particles that are \textit{isotropically} distributed:
\begin{equation}
    \mu(\textbf{p}_{0}) d^{3}\textbf{p}_{0} =\frac{1}{4 \pi}  S(p_{0}) dp_{0}\;d\Omega_{0},
    \label{isotropic}
\end{equation}
where $\int dp_{0} \; S(p_{0})=1$ (such that $\int d^{3} \textbf{p}_{0} \; \mu(\textbf{p}_{0})=1$). 
Therefore, we obtain
\begin{equation}
    \begin{aligned} \frac{d \rho_{S}}{dt}(\textbf{x}, &\textbf{y},t) = -2(2\pi)^{\frac{3}{2}}  \int dp_{0} S(p_{0}) \;nv(p_{0}) \int \frac{d \Omega_{0} d\Omega '}{4 \pi} \\
    &\times \frac{d \sigma}{d \Omega '} \left\{-e^{-ip_{0}\left(\hat{n}'-\hat{n}_{0}\right)\cdot(\textbf{x}-\textbf{y})} +1\right\}\rho_{S} (\textbf{x}, \textbf{y},t) ,\end{aligned}
    \label{manypart}
\end{equation}
where we have not written explicitly the spin labels $r_{i}$ and $r_{f}$ for brevity. Hence, only considered the spatial part of $\hat{\rho}_{S}$. As can be seen from Eq.(\ref{manypart}), the time evolution of the density matrix will be of the type:
\begin{equation}
    \begin{split}
        \frac{d \rho_{S}(\textbf{x},\textbf{y},t)}{dt} &= - \Gamma \cdot \rho_{S}(\textbf{x},\textbf{y},t) \\
        \Rightarrow \rho_{S}(\textbf{x},\textbf{y},t)&=e^{- \Gamma t}\rho_{S}(\textbf{x},\textbf{y},0), \label{eq:dens_mat_evol}
        \end{split}
\end{equation}
with 
%
\begin{equation}
    \begin{aligned} 
    \Gamma \equiv 2(2\pi)^{3/2}  \int dp_{0} S(p_{0}) \;nv(p_{0}) \int \frac{d \Omega_{0} d\Omega '}{4 \pi} \\
    \times \frac{d \sigma}{d \Omega '}(\hat{n}_{0},\hat{n}^{'}) \left\{-e^{-ip_{0}\left(\hat{n}'-\hat{n}_{0}\right)\cdot(\textbf{x}-\textbf{y})}+1\right\} .\end{aligned}
    \label{Gamma}
\end{equation}
%
Eq.~\eqref{eq:dens_mat_evol} shows the role played by $\Gamma$: it suppresses the off-diagonal density matrix elements, while leaving the diagonal matrix elements unchanged. The decay factor $\Gamma$ dictates the decoherence rate~\cite{Schlosshauer:2014pgr}.

In the limit $t \rightarrow + \infty$, these off-diagonal elements are completely suppressed, i.e. they become asymptotically $0$.
This means that the density matrix $\rho_{S}$ becomes diagonal, indicating a classical mixed state. 
The environment thus over time acquires information from the system, leading to the information leaking to the environment, requiring an observer to measure both the crystal and environment to regain the system's quantum information.
This is why $\Gamma$ as given in Eq.~\eqref{Gamma} is called the decoherence rate, it causes the decoherence of the system over time.

\section{Dipole-Dipole interaction }\label{DD}

In this section, we will consider a special case of electromagnetic interaction; the dipole-dipole interaction. This type of interaction is relevant also in the case of neutral test masses. Typically crystals such as diamond possesses dielectric properties. Therefore, we will consider a neutral diamond type crystal. Diamnond also provides defects such as NV centers, which helps to create the spatial superposition, see~\cite{Bose:2017nin,PhysRevLett.125.023602,Marshman:2021wyk,Zhou:2022epb}. 
In this section, we have also converted the expressions from the previous section to the SI units, as we will discuss the physics of the experimental aspects.

To obtain an explicit expression for the decoherence rate (see Eq.~\eqref{Gamma}), we have to specify the differential cross section related to the type of interaction between the environment and the crystal. 
Therefore, let us assume that the environmental particles are gas molecules with a non-zero dipole moment. The dipole-dipole potential is given by~\cite{jackson_classical_1999}:
%
\begin{equation}
\begin{split}
    V(\textbf{r})= \frac{1}{4 \pi \epsilon_{0}} \frac{\textbf{d}_{1} \cdot \textbf{d}_{2} - 3 \left( \textbf{n} \cdot \textbf{d}_{1} \right) \left( \textbf{n} \cdot \textbf{d}_{2}  \right)}{|\textbf{r}|^3},
\end{split}
    \label{potential_complete}
\end{equation}
\begin{figure}[t]
\centering
\includegraphics[scale=0.45]{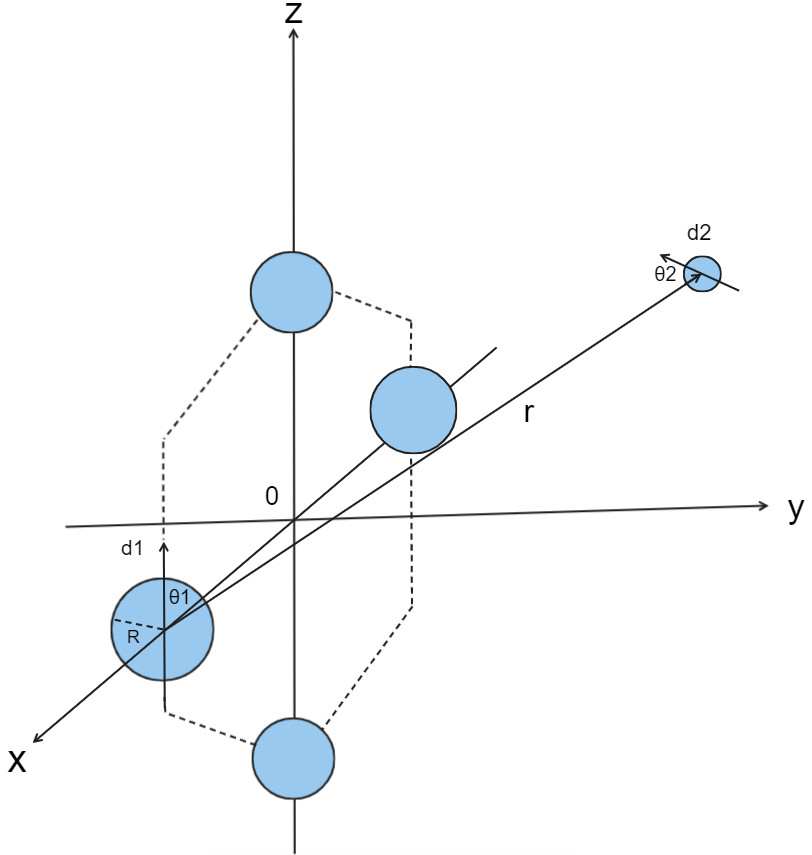}
\caption{The figure shows a schematic representation of the dipole-dipole interaction. The crystal has a finite size $R$ and possesses a dipole $\textbf{d}_{1}$ parallel to the $\hat{\textbf{z}}$-axis, while the environmental particle has a dipole $\textbf{d}_{2}$ which is randomly oriented. The vector $\textbf{r}$ denotes the distance between the centers of the two dipoles, while we can define the orientation of the two dipoles, $\textbf{d}_{1}$ and $\textbf{d}_{2}$, with respect to this vector as $\theta_{1}$ and $\theta_{2}$, respectively. The time evolution of the superposition size is illustrated by the dotted lines.}
\label{drawing}
\end{figure}
where $\textbf{d}_{1}$ and $\textbf{d}_{2}$ are the electric dipole of the crystal and of the environmental particle respectively, $\textbf{r}$ represents the distance between the centers of the two dipoles, and $\textbf{n}=\textbf{r}/|\textbf{r}|$ is its associated unit vector. Figure \ref{drawing} shows a schematic representation of a generic dipole-dipole interaction, where the crystal is represented to be in a spatial superposition. We have also introduced the angles $\theta_{1}$ and $\theta_{2}$ between $\textbf{r}$ and respectively $\textbf{d}_{1}$ and $\textbf{d}_{2}$.

Our goal is to compute the differential cross-section for this type of potential. A very useful tool for this purpose is the Born approximation
, which expresses the differential cross section ${d \sigma}/{d \Omega}$ in terms of the Fourier Transform $\Tilde{V}(q)$ of the potential $V(\textbf{r})$:
\begin{equation}
    \frac{d \sigma}{d \Omega}=\frac{m^2}{4 \pi^2 \hbar^4} |\Tilde{V}(q)|^2,
    \label{Born_approx_formula}
\end{equation}
with $\Tilde{V}(q)$ is the Fourier transform of the potential given by Eq.~\eqref{potential_complete}, i.e. $\Tilde{V}(\textbf{q})=\int d^3 \textbf{x} \;e^{\frac{i}{\hbar} \textbf{q} \cdot \textbf{x}} V(\textbf{r})$. In particular, defining the vectorial components for $\textbf{d}_{1}=(0,0,d_{1})$, $\textbf{d}_{2}=(d_{2x},d_{2y},d_{2z})$ and $\textbf{n}=(\sin(\Bar{\theta})\cos(\Bar{\phi}),\sin(\Bar{\theta})\sin(\Bar{\phi}),\cos(\Bar{\theta}))$, we can perform the angular integral over $d \Bar{\Omega}=d\Bar{\phi} d\Bar{\theta} \; \sin(\Bar{\theta})$:
\begin{equation}
\begin{split}
   &\int d \Bar{\phi} d\Bar{\theta} \; \sin(\Bar{\theta}) \frac{e^{\frac{i}{\hbar} q r \; \cos(\Bar{\theta})}}{4 \pi \epsilon_{0}} \frac{\textbf{d}_{1} \cdot \textbf{d}_{2} - 3 \left( \textbf{n} \cdot \textbf{d}_{1} \right) \left( \textbf{n} \cdot \textbf{d}_{2}  \right)}{r^3} \\
   &= -\frac{2d_{1}d_{2z}}{\epsilon_{0}} \frac{\hbar^3}{{q^3 r^6}} \bigg[ \left( \frac{r^2}{\hbar^2}q^2-3 \right) \sin\big( \frac{r}{\hbar} q \big) \\ 
   &\qq{}\qq{}\qq{}\qq{}\qq{}+ 3 \frac{r}{\hbar} q \; \cos\big( \frac{r}{\hbar} q \big) \bigg]
\end{split} \label{only_angular_FT}
\end{equation}
where $\textbf{q}=\textbf{p}_{f}-\textbf{p}_{0}$ is the transferred momentum of the scattered particle.

In a low-energy regime that we have assumed, the environmental particle will not be able to penetrate the crystal. We therefore assume a minimal distance of interaction equal to the radius of the crystal, $R$. This gives a lower limit for the integral over $r$:
\begin{equation}
    \begin{split}
        \Tilde{V}(q)&=-\frac{2d_{1}d_{2z} \hbar^3}{\epsilon_{0} q^3} \int_{R}^{+\infty}dr\;r^2 \\
        &\qq{}\times \frac{1}{{r^6}} \bigg[ \left( \frac{r^2}{\hbar^2}q^2-3 \right) \sin\big( \frac{r}{\hbar} q \big) + 3 \frac{r}{\hbar} q \; \cos\big( \frac{r}{\hbar} q \big) \bigg]\\
        &= \frac{2 d_{1} d_{2z} \hbar^3}{\epsilon_{0} R^3 q^3} \bigg[ \sin\big( \frac{R}{\hbar} q \big) - \frac{R}{\hbar} q\; \cos\big( \frac{R}{\hbar}q \big) \bigg].
    \end{split}
\label{FT_analytic}
\end{equation}
The minimum interaction distance $R$ in this case is completely equivalent to considering a form factor $F(q)$, which is compulsory especially when the environmental particle is able to resolve the dimensions of the crystal, i.e. when $\lambda \ll R$. 
In particular, considering a constant charge distribution density for the crystal $\rho(\textbf{x}) \equiv c$ for $r \leq R$ and $\rho(\textbf{x}) \equiv 0$ otherwise, the form factor will be given by:

\begin{align}
        F(q)&= \int d^3 \textbf{x} \; e^{\frac{i}{\hbar} \textbf{q} \cdot \textbf{x}} \rho(\textbf{x}) = \int d\Omega \int_{0}^{R} dr \; r^2 e^{\frac{i}{\hbar} \textbf{q} \cdot \textbf{x}} \cdot c \nonumber \\
        &\propto \frac{ \sin\big( \frac{R}{\hbar} q \big) - \frac{R}{\hbar} q\; \cos\big( \frac{R}{\hbar}q \big) }{\big( \frac{R}{\hbar} q \big)^3} \, ,
\end{align}
which is exactly the term that appears inside the parentheses of the right-hand-side of Eq.~\eqref{FT_analytic}.

Eq.~\eqref{FT_analytic} can be inserted inside the Born formula in order to find the differential cross section ${d \sigma}/{d \Omega}$. 
After computing the modulus squared of $\tilde{V}(\boldsymbol{q})$ (see Eq.~\eqref{FT_analytic}) we can average the final result over the orientation of the environmental dipole $d_{2}$, which in Eq.~(\ref{FT_analytic}) appears with the term $d_{2z}=d_{2} \cos(\theta_{2})$. We can schematically summarize these operations as: $\Tilde{V}(q)\rightarrow |\Tilde{V}(q)|^2 \rightarrow \int d \phi_{2} d \theta_{2} \; \sin(\theta_{2}) |\Tilde{V}(q)|^2$.

The final expression for the differential cross-section will thus be given by:
\begin{equation}
\begin{split}
     \frac{d \sigma}{d \Omega}= \frac{4 \hbar^2 m^2 d_{1}^2 d_{2}^2}{3 \pi \epsilon_{0}^2 R^6 q^6} \bigg[ \sin\big( \frac{R}{\hbar} q \big) - \frac{R}{\hbar} q\; \cos\big( \frac{R}{\hbar}q \big) \bigg]^2,
\end{split}   \label{diffcsdipdip}
\end{equation}
where $m$ is the mass of the environmental dipole, $d_1$ ($d_2$) is the strength of the crystal (environmental) dipole, $R$ is the radius of the crystal, and $q$ the momentum transfer from the scattered particle to the crystal. $\hbar$ is the Planck constant and $\epsilon_0$ is the vacuum permittivity.
We use Eq.~\eqref{diffcsdipdip} to find the decoherence rate in the long-wavelength and short-wavelength approximations.

\subsection{Short-wavelength limit approximation}\label{sec_short}

In the short-wavelength limit the wavelength of the environmental particle, $\lambda_0$, is small compared to the superposition size, $\Delta x$, of the spatially superposed crystal, see Fig.\ref{drawing}:
$$\lambda_{0} \ll \Delta x \, .$$ 
This implies that $p_{0} \Delta x \gg 1$, meaning that when the phase exponential $e^{-ip_{0}(\hat{n}'-\hat{n}_{0})\cdot (\Vec{x}-\Vec{y})}$ in Eq.~\eqref{Gamma} is integrated over $p_{0}$, it will oscillate very fast and its contribution will be negligible.
Therefore the approximation of Eq.~\eqref{Gamma} for the decoherence rate in the short-wavelength limit becomes:
\begin{equation}
    \begin{aligned} 
    \Gamma_{S}=2(2\pi)^{3/2}  \int dp_{0} S(p_{0}) \;nv(p_{0})\;\sigma_{\text{CM}}(p_0) ,
    \end{aligned}
    \label{Gammashort}
\end{equation}
where 
$$\sigma_{\text{CM}}=\int \frac{d \Omega_{0} d\Omega '}{4 \pi} \frac{d \sigma}{d \Omega '}(\hat{n}_{0},\hat{n}^{'}),$$ 
is the total cross-section and the subscript $S$ in $\Gamma_{S}$ stands for `Short', indicating that we are computing the decoherence rate in the short-wavelength limit.

Before computing explicitly $\sigma_{\text{CM}}$, let us find out where the angular dependence comes from inside Eq.~(\ref{diffcsdipdip}). We know that $q$ is the transferred momentum: $\Vec{q}=\Vec{p'}_{f}-\Vec{p}_{0}$. However, we have assumed that the environmental particle is significantly lighter than the crystal and that the environmental dipole possesses lower energy compared to the mass-energy of the crystal. These approximations were already introduced in the Section~\ref{sec:step2_deco}. 
In particular, as can be seen from Eq.~(\ref{twentysix}), these assumptions lead to the conservation of energy through Dirac's delta $\delta (E_{p'_{f}}-E_{p_{0}})$. This means that $E_{p'_{f}}=E_{p_{0}}$ and therefore $p'_{f}=p_{0}$, i.e. the \textit{modulus} of the momentum of the environmental particle is the same before and after the scattering process. Only the \textit{direction} of the final momentum is different from the initial one, given our assumptions.

This means that the transferred momentum $q$ will have the following expression:
\begin{equation}
    \begin{split}
        q
        &=\sqrt{p_{f}^{\prime 2}+p_{0}^2-2 \Vec{p}\;'_{f} \cdot \Vec{p}_{0}}=\sqrt{2p_{0}^2-2 p_{0}^2 \cos(\theta')}\\
        &= 2 p_{0} \sin(\theta'/2).
    \end{split} \label{transf_moemntum}
\end{equation}

Using Eqs.~(\ref{diffcsdipdip}),~\eqref{transf_moemntum} we compute $\sigma_{\text{CM}}$:
\begin{widetext}
\begin{equation}
\begin{split}
    \sigma_{\text{CM}}&=\int \frac{d \Omega_{0}}{4 \pi} \int d \Omega '  \frac{4 \hbar^2 m^2 d_{1}^2 d_{2}^2}{3 \pi \epsilon_{0}^2 R^6 q^6} \bigg( \sin\Big( \frac{R}{\hbar} q \Big) - \frac{R}{\hbar} q\; \cos\Big( \frac{R}{\hbar}q \Big) \bigg)^2\\
    &= \frac{\hbar^2 m^2 d_{1}^2 d_{2}^2}{48 \pi \epsilon_{0}^2 R^6 p_{0}^6} 2 \pi \int_{0}^{\pi} d\theta'\; \sin(\theta')\frac{1}{\sin^{6}(\theta'/2)} \bigg( \sin\Big( 2\frac{R}{\hbar} p_{0} \sin(\theta'/2) \Big) - 2 \frac{R}{\hbar} p_{0} \sin(\theta'/2) \; \cos\Big( 2 \frac{R}{\hbar}p_{0} \sin(\theta'/2) \Big) \bigg)^2\\
    &= \frac{m^2 d_{1}^2 d_{2}^2 \hbar^2}{48 \epsilon_{0}^2 R^6 p_{0}^{6}} \bigg[ -1-8 \left(  \frac{R}{\hbar} p_{0} \right)^{2} +32 \left(  \frac{R}{\hbar} p_{0} \right)^{4}+ \cos\left( 4 \frac{R}{\hbar} p_{0} \right) +4 \frac{R}{\hbar} p_{0} \; \sin\left( 4 \frac{R}{\hbar} p_{0} \right) \bigg].
\end{split} \label{totcsquasi}
\end{equation}    
\end{widetext}

In order to compute $\Gamma_{S}$ using Eq.(\ref{Gammashort}) we choose a distribution $S(p_{0})$ for the environmental particles.
Supposing that the environmental dipoles form a thermal bath with temperature $T$, the average energy of one environmental particle will be $\Bar{E} \sim k_{B} T$, which leads to an average momentum of $\Bar{p} \sim \sqrt{2 m k_{B}T}$. Therefore, we consider for $S(p_{0})$ the Maxwell-Boltzmann distribution:
\begin{equation}
    \begin{aligned}
        S(p_{0})=4 \pi p_{0}^2 \Big( 2 \pi m k_{B} T \Big)^{-3/2} e^{- \frac{p_{0}^2}{2 m k_{B} T}},
    \end{aligned}
    \label{Max_Boltz}
\end{equation}
which satisfies the requirement $\int dp_{0}\;S(p_{0})=1$. 
For very low temperatures Eq.~(\ref{Max_Boltz}) is has a strong peak around its mean value $\Bar{p}=\sqrt{2 m k_{B} T}$. This is because Eq.~(\ref{Max_Boltz}) is a Gaussian function of the type $S(p_{0}) \sim e^{-{p_{0}^2}/{\sigma}}$, where $\sigma$ represents the standard deviation, which intuitively tells us how much the distribution is spread over the momentum space. 
In our case we have $\sigma \propto m k_{B} T$, where $m$ is the environmental particle's mass which can be considered to be molecules or atoms or air molecules. 
Therefore, at low temperatures (i.e. $T \sim 1\;\text{K}$) we will have that $\sigma \ll 1$ and the Maxwell-Boltzmann function will be a very narrow distribution centered around $\Bar{p}=\sqrt{2 m k_{B} T}$.

We can thus consider all the particles to have approximately the same momentum $\Bar{p}$, i.e. $S(p_{0})=\delta(p_{0}-\Bar{p})$. 

Putting the expression for the total cross section given by Eq.~(\ref{totcsquasi}) inside Eq.~(\ref{Gammashort}), we find the decoherence rate due to dipole-dipole interactions in the short wavelength limit:
\begin{widetext}
\begin{equation}
    \begin{split}
        \Gamma_{S} &= 2(2 \pi)^{\frac{3}{2}} \int dp_{0}\; \delta(p_{0}-\Bar{p}) n \frac{p_{0}}{m}\frac{m^2 d_{1}^2 d_{2}^2 \hbar^2}{48 \epsilon_{0}^2 R^6 p_{0}^{6}}  \bigg[ -1-8 \left(  \frac{R}{\hbar} p_{0} \right)^{2} +32 \left(  \frac{R}{\hbar} p_{0} \right)^{4}+ \cos\left( 4 \frac{R}{\hbar} p_{0} \right) +
        4 \frac{R}{\hbar} p_{0} \; \sin\left( 4 \frac{R}{\hbar} p_{0} \right) \bigg] \\
        &=(2 \pi)^{\frac{3}{2}}\frac{\hbar^2 m d_{1}^2 d_{2}^2 n}{24 \epsilon_{0}^2 R^6 \Bar{p}^{5}} \bigg[ -1-8 \left(  \frac{R}{\hbar} \Bar{p} \right)^{2} +32 \left(  \frac{R}{\hbar} \Bar{p} \right)^{4} + \cos\left( 4 \frac{R}{\hbar} \Bar{p} \right) +4 \frac{R}{\hbar} \Bar{p} \; \sin\left( 4 \frac{R}{\hbar} \Bar{p} \right) \bigg].
    \end{split} \label{final_dec_long}
\end{equation}    
\end{widetext}
Here, $m$ is the mass of the environmental dipole with dipole strength $d_2$, while $d_1$ is the strength of the crystal's dipole, $R$ is the radius of the crystal, $\bar{p}$ denotes the mean momentum magnitude $\Bar{p}=\sqrt{2 m k_{B} T}$ and $n$ is the environmental particle's number density, i.e. $\big[ n \big]=\text{m}^{-3}$. $\hbar$ is the Planck constant and $\epsilon_0$ is the vacuum permittivity.


\subsection{Long-wavelength approximation}\label{sec_long}
We will now consider the long-wavelength approximation to Eq.~\eqref{Gamma}, where the wavelength associated with the environmental particle is much bigger than the superposition size, see Fig.~\ref{drawing} 
$$\lambda_{0} \gg \Delta x \, .$$
As a result, in the long-wavelength approximation $ip_{0}(\hat{n}'-\hat{n}_{0})\cdot (\Vec{x}-\Vec{y}) \ll 1$, and we can take the Taylor expansion of the exponential that appears in the right-hand-side of Eq.~(\ref{Gamma}) to obtain \cite{Schlosshauer:2014pgr}:
\begin{equation}
    \frac{i}{\hbar}p_{0}\left(\hat{n}'-\hat{n}_{0}\right)\cdot(\Vec{x}-\Vec{y})+\frac{1}{2 \hbar^2}p_{0}^2[\left(\hat{n}'-\hat{n}_{0}\right)\cdot(\Vec{x}-\Vec{y})]^2.
\end{equation}
The first term  in the equation above gives an integral of an odd function, due to the fact that $\hat{n}'-\hat{n}_{0}$ is antisymmetric in the exchange of $\hat{n}$ and $\hat{n}'$, while $\frac{d \sigma}{d \Omega} (\hat{n}_{0},\hat{n}')$ is symmetric, giving a total odd function. 

The second term can be simplified by assuming that the particular direction $\Vec{x}-\Vec{y}=|\Vec{x}-\Vec{y}|\;\hat{s}=\Delta x \; \hat{s}$ of the scattering center (i.e. of the crystal) does not depend on the direction $\hat{s}$.
We can thus average this term over all possible directions $\hat{s}$, obtaining \cite{Schlosshauer:2014pgr}:
\begin{equation}
\begin{split}
    \left( \Delta x \right)^2 \frac{1}{3} \sum_{s=x,y,z} \left[\hat{s} \cdot \left( \hat{n}'-\hat{n}_{0} \right) \right]^2=\frac{1}{3} \left( \Delta x \right)^2 |\hat{n}'-\hat{n}_{0}|^2\\
    =\frac{2}{3} \left( \Delta x \right)^2 |1-\hat{n}' \cdot \hat{n}_{0}|=\frac{2}{3} \left( \Delta x \right)^2 (1-\cos(\theta ')).
\end{split}
\end{equation}
where $\theta'$ is the scattering angle.\\
Performing the angular integral in Eq.~\eqref{Gamma} gives:
\begin{equation}
\begin{split}
    &\int d\Omega ' \frac{d \sigma}{d \Omega '}(\hat{n}_{0},\hat{n}^{'})\frac{2}{3}  (1-\cos(\theta '))\\
    &\qq{}=\frac{2 \pi}{3} \int d(\cos(\theta ')) (1-\cos(\theta ')) \frac{d \sigma}{d \Omega '}(\hat{n}_{0},\hat{n}^{'}) \equiv \sigma_{\text{eff}} ,    
    \label{sigma_eff_generic}
\end{split}
\end{equation}
where we have integrated over the azimutal angle $\int_{0}^{2 \pi} d \phi =2 \pi$ and we have defined the effective cross section $\sigma_{\text{eff}}$. 
Using Eq.~\eqref{Gamma}, the final expression for the decoherence rate $\Gamma$ in the long-wavelength limit is:
\begin{equation}
    \Gamma_{L}= 2(2 \pi)^{3/2} \Delta x^2 \int dp_{0}\; S(p_{0}) n \frac{p_{0}}{m} \sigma_{\text{eff}}(p_{0}) \frac{p_{0}^2}{\hbar^2}.
    \label{Dec_rate_longWL}
\end{equation}
which is an expression that can also be found in the literature \cite{Schlosshauer:2014pgr,joos2003decoherence}. 
The subscript $L$ stands for `Long', since we have computed the decoherence rate in the long-wavelength limit.

Comparing Eq.~(\ref{Dec_rate_longWL}) and Eq.~(\ref{Gammashort}) we notice that we can obtain the long-wavelength limit expression by substituting $\sigma_{\text{CM}} \longrightarrow \sigma_{\text{eff}}$ inside the short-wavelength limit formula and by multiplying it by the term $\Delta x^2 {p_{0}^2}/{\hbar^2}$. 

In particular, $\sigma_{\text{eff}}$ will be of the same order as $\sigma_{\text{CM}}$, given that the only difference between the two is a purely geometrical factor $\left( 1-\cos(\theta ') \right)$ inside Eq.~(\ref{sigma_eff_generic}). Therefore, the main difference between the two expressions is entirely encoded inside the term $\Delta x^2 {p_{0}^2}/{\hbar^2}$.
When the \textit{physical} situation is that of the short-wavelength limit ($\lambda_{0} \sim {\hbar}/{p_{0}} \ll \Delta x$), then  $\Delta x^2 {p_{0}^2}/{\hbar^2} \gg 1$, and the following inequality holds:
\begin{equation}
    \frac{\Gamma_{S}}{\Gamma_{L}} \sim \frac{\hbar^2}{\Delta x^2 p_{0}^2} \ll 1 \, \Rightarrow \, \Gamma_{S} \ll \Gamma_{L}.
    \label{comparison_L_S}
\end{equation}
This means that when the physical limit is the short-wavelength limit, the long-wavelength decoherence rate given in Eq.~(\ref{Dec_rate_longWL}) can be used as an {\it upper bound} estimation of the decoherence rate.

We now compute $\sigma_{\text{eff}}$ for the dipole-dipole interaction case. Plugging Eq.~(\ref{diffcsdipdip}) into Eq.~(\ref{sigma_eff_generic}), we obtain:
\begin{equation}
    \begin{split}
        \sigma_{\text{eff}} &= \frac{2}{3} \pi \int_{-1}^{1} d\big[ \cos(\theta ') \big] \big(  1-\cos(\theta ')\big) \\
        &\qq{}\times \frac{4 \hbar^2 m^2 d_{1}^2 d_{2}^2}{3 \pi \epsilon_{0}^2 R^6 q^6} \bigg( \sin\big( \frac{R}{\hbar} q \big) - \frac{R}{\hbar} q\; \cos\big( \frac{R}{\hbar}q \big) \bigg)^2 \\
        &= \frac{2}{3} \pi \beta \int_{0}^{1} \dd{u} \; 4 u \cdot 2u^2 \frac{\big[  \sin\big(a u \big) - a u\; \cos\big( a u \big) \big]^2}{u^6},
    \end{split}
    \label{cs_eff_tot}
\end{equation}
where in the very last line we have substituted $q$ with $q= 2 p_{0} \sin(\theta '/2)$ as in Eq.~(\ref{transf_moemntum}), and we have defined
\begin{equation}
    \beta \equiv \frac{ \hbar^2 m^2 d_{1}^2 d_{2}^2}{48 \pi \epsilon_{0}^2 R^6 p_{0}^6} \, ,\qq{} a \equiv 2 {R p_0}/{\hbar}
\end{equation}
Moreover, we defined $u \equiv \sin(\theta '/2)$ and used $\cos(\theta')=1-2\; \sin(\theta '/2) \equiv 1-2u^2$, to rewrite the integral:
\begin{equation}
    \begin{split}
        \int_{-1}^{1} d\big[ \cos(\theta ') \big] \big(  1-\cos(\theta ')\big)
        &=\int_{0}^{1} \dd{u} \; 4u \cdot 2u^2.
    \end{split}
\end{equation}
The integral inside Eq.~(\ref{cs_eff_tot}) can now be computed analytically:
\begin{align}
        \sigma_{\text{eff}}&=\frac{16}{3} \pi \beta \int_{0}^{1} \dd{u} \frac{\big[  \sin\big(a u \big) - a u\; \cos\big( a u \big) \big]^2}{u^3}\\
   &= \frac{4}{3} \pi \beta \bigg[ a^2\big( \ln(4 a^2) -2\; \text{Ci}(2a)+2(\gamma-1) \big) \nonumber\\
       &\qq{}\qq{}\qq{} + 2a\; \sin(2a) + \cos(2a) -1 \bigg],
    \label{new_sigma-Eff}
\end{align}
where $\gamma \simeq 0.577$ is the Euler-Mascheroni constant and $$\text{Ci}(2a)=-\int_{2a}^{+\infty} dt \frac{\cos(t)}{t} \, ,$$ is a known trigonometric integral.

In the limit $p_{0} \to 0$ (i.e. $a \to 0$) the effective cross section (Eq.~\eqref{new_sigma-Eff}) is well-defined, it goes to zero~\footnote{This can be easily seen from the expansion of $\text{Ci}(x)=\gamma+\ln(x)+\sum_{n=1}^{+\infty}\frac{(-x^2)^n}{2n(2n)!}=\gamma+\ln(x)-\frac{x^2}{2 \cdot 2!}+\frac{x^4}{4 \cdot 4!}+...$ For $x \rightarrow 0$, one can see that the only terms in the expansion that survive are $\gamma+\ln(x)$, i.e. for small $x$ we have that $\text{Ci}(x)-\ln(x) \sim \gamma$, which is the reason why the overall limit is $0$.}.
Since $p_{0}\propto T$, this means that the decoherence rate is well-defined for very small temperatures. 
Specifically in the limit $T\to0$, the decoherence rate given in Eq.~\eqref{Dec_rate_longWL} becomes $\Gamma_L\to 0$.

We compute explicitly Eq.(\ref{Dec_rate_longWL}) in terms of the newly defined effective cross section. 
We can again consider a momentum distribution of the environmental particles like $S(p_{0})=\delta(p_{0}-\Bar{p})$, and further approximate Eq.~(\ref{new_sigma-Eff}) using the property $\lim_{z \to \infty}\text{Ci}(z)=0$. 
This is possible when, as can be seen from Eq.~(\ref{new_sigma-Eff}), ${R \Bar{p}}/{\hbar} \gg 1$, true in most experimental setups \cite{Bose:2017nin,Afek:2021bua, Carney_2019,Barker:2022mdz}. 
The average momentum from atomic masses at low temperatures ($\sim 1\ \text{ K}$) is $\Bar{p}=\sqrt{2 m k_{B} T} \sim 10^{-25} \text{kg}\; \text{m} \; \text{s}^{-1}$.

For a micron size spherical crystal one thus finds ${R \Bar{p}}/{\hbar} \sim 10^3 \gg 1$, meaning that $\text{Ci}(10^3) \sim 10^{-4}$.
We therefore approximate $\sigma_{\text{eff}}$ as (see Eq.~(\ref{new_sigma-Eff})):
\begin{equation}
    \begin{split}
        \sigma_{\text{eff}}(\Bar{p}) \simeq \frac{4}{3} \pi \beta a^2 \ln(4 a^2)
        = \frac{2 m^2 d_{1}^2 d_{2}^2}{9 \epsilon_{0}^2 R^4 \Bar{p}^4} \ln\bigg( 4 \frac{R}{\hbar} \Bar{p} \bigg).
    \end{split}
\end{equation}
Using this expression for $\sigma_{\text{eff}}(p)$ inside Eq.~\eqref{Dec_rate_longWL} we obtain:
\begin{equation}
    \Gamma_{L}= (2 \pi)^{\frac{3}{2}} \frac{4 m d_{1}^2 d_{2}^2 n \Delta x^2}{9 \epsilon_{0}^2 \hbar^2 R^4 \Bar{p}} \ln\bigg( 4 \frac{R}{\hbar} \Bar{p} \bigg),
    \label{Gam}
\end{equation}
which represents the final formula for the decoherence rate due to dipole-dipole interactions in the long-wavelength limit.

As mentioned before, $m$ is the mass of the environmental dipole, $d_1$ ($d_2$) is the strength of the crystal (environmental) dipole, $R$ is the radius of the crystal, $\bar{p}$ denotes the mean momentum magnitude $\Bar{p}=\sqrt{2 m k_{B} T}$, $n$ is the environmental particle's number density, i.e. $\big[ n \big]=\text{m}^{-3}$, $\hbar$ is the Planck constant and $\epsilon_0$ is the vacuum permittivity. 
Compared to Eq.~(\ref{final_dec_long}), Eq.~\eqref{Gam} is dependent on the spatial superposition size of the crystal, $\Delta x$.

\section{QGEM experiment}\label{sec:QGEM}

In this section, we explore the decoherence of the dipole-dipole interaction in the context of the QGEM experiment.
Although different setups have been proposed, generally in the QGEM experiment~\cite{Bose:2017nin}, a spatial superposition is created of size $\Delta x$ and kept adjacent to each other  for a time $\tau \sim 1\;\text{s}$. 
In the QGEM experiment the neutral test masses, which we  take to be diamond micron size crystals with an embedded spin in their NV centre, interact only via gravity, see for details~\cite{Bose:2017nin,vandeKamp:2020rqh}.
The quantum nature of gravity entangles the test masses, and therefore the quantum nature of gravity can be empirically measured by witnessing the generation of an entangled state from an unentangled state~\cite{Bose:2017nin,Marshman:2018upe,Bose:2022uxe,Danielson:2021egj,Carney_2019,Christodoulou:2022vte}. 
It is very important to keep track of all the possible sources of decoherence, i.e. everything that can lead to a destruction of the spatial superposition. 
Considering that the superpositions are held for a time $\tau \sim 1 \; \text{s}$, we require the decoherence rate to not be bigger than $\Gamma_{S}= 10^{-2} \text{ Hz}$. Such a value for $\Gamma_{S}$ leads to a decoherence time $\tau_d = 10^{2} \text{ s} \gg \tau$, therefore making sure that the coherence loss due to the interaction with the environment is small during the realization of the experiment. This value of the decoherence rate has been deemed to be safe by various other considerations, see~\cite{vandeKamp:2020rqh,Toros:2020dbf,Tilly:2021qef,Schut:2021svd}.
For a diamond density of $\rho = 3.5 \times 10^{3} \text{kg}/\text{m}^3$, and for the mass of $\sim 10^{-14} \text{ kg}$ is required to witness the entanglement in the QGEM experiment~\cite{Bose:2017nin}. We also assume that the experimental box with sides of length $L\sim {\cal O}(1)$cm~\cite{Toros:2020dbf}.

One of the unavoidable sources of decoherence during the experiment arises from electromagnetic interactions. Specifically, despite creating a vacuum with extremely low pressure inside the experimental box \cite{Toros:2020dbf}, there is still a possibility that some random air molecules may inadvertently persist within the box. Such molecules are neutral  but can possess an electric dipole $d_{2}$ of the order of $d_{2} \sim 1\;\text{D}$  \cite{table}. This means that the interaction between an environmental particle and the crystal in superposition is possible if the crystal is made of a dielectric medium and/or if the crystal also has a dipole $d_{1}$ too, see \cite{abdul,Afek:2021bua}.

In this section we are going to analyze three different physical situations that are likely to happen in a QGEM setup.

\begin{itemize}

\item In the first case, (subsection~\ref{subsec_induced_crystal}) the environmental particles generate an electric field inside the crystal, which will induce a dipole moment in the crystal because of its dielectric properties. 
\item In the second case (subsection~\ref{subsec_d1_d2}), the diamond is considered to have a permanent dipole, as estimated by some experiments \cite{Afek:2021bua}. This permanent dipole moment $d_{1}$ it is treated as a free parameter, on which we put constraints by requiring that the decoherence rate has to be smaller than $\Gamma_{S} \sim 10^{-2} \text{Hz}$. 
\item Finally, in the last case (subsection~\ref{subsec_induced_environment}) the crystal is considered to have a permanent dipole, which will generate an electric field that induces an electric dipole moment in the environmental particles. Considering neutral air molecules (e.g. $\text{N}_{2}, \text{O}_{2}, \text{Ar}$ and $ \text{CO}_{2}$) we will be able to find their dipole moment $d_{2}$ induced by the crystal and we will analyze the resulting decoherence rate.

\end{itemize}


\subsection{Crystal's dipole induced by the environment}\label{subsec_induced_crystal}

The first case that we  will analyze is when $d_{1}$ is induced by the environment. In fact, being a dielectric material, the diamond crystal has a polarizability $\alpha$ when subjected to an external electric field $E_{\text{ext}}$, which inside the crystal is perceived as a local field $E_{\text{loc}}=E_{\text{ext}}/\epsilon_{r}$, with $\epsilon_{r} \sim 5.7$ being the relative dielectric constant of the crystal.
In particular, in isotropic media, the external field will create a local dipole in each atom of the crystal's lattice \cite{jackson_classical_1999}:
\begin{equation}
    d_{1}=\alpha \, E_{\text{loc}}.
    \label{induced_63}
\end{equation}
This means that the total contribution will be given by the sum over all $N'$ atoms of the lattice:
\begin{equation}
    d_{1}=N' \alpha E_{\text{loc}}.
    \label{N'alphaEloc}
\end{equation}
So in order to find an explicit expression for $d_{1}$, we need the expressions for $\alpha$ and $E_{\text{loc}}$. For $\alpha$ we have the Classius-Mossotti relation:
\begin{equation}
    \frac{n' \alpha}{3 \epsilon_{0}}=\frac{\epsilon_{r}-1}{\epsilon_{r}+2},
    \label{Classius-Mossotti}
\end{equation}
where $n'=\frac{N'}{V}$ is the atomic density inside the crystal, the volume is given by $V = \frac{4\pi}{3} R^3$ if we assume the crystals to be perfect spheres.
The induced electric field will be generated by the environmental dipole $d_{2}$ \cite{jackson_classical_1999}:
\begin{equation}
    E_{\text{loc}} = \frac{d_{2}}{2 \pi \epsilon_{0} \epsilon_{r}} \frac{1}{\Bar{r}^3},
    \label{E_loc}
\end{equation}
where we have considered the angular dependence inside Eq.~(\ref{potential_complete}) to be maximum, e.g. $\theta_{1}=0$ and $\theta_{2}= \pi$, which results in a factor of $2$ from the squared parenthesis of Eq.~(\ref{potential_complete}). 
Here, $\Bar{r}$ represents the average interaction distance between an environmental particle 
and the crystal. In the worst case scenario, this average distance is given by $\Bar{r} \simeq R \sim 10^{-6} \text{m}$
, which is the closest possible distance between $d_{1}$ and $d_{2}$. 
We consider the environment to be composed of atomic dipoles, for example by Helium atoms. This means that we can consider $d_{2}$ to be \cite{table}:
\begin{equation}
\begin{aligned}
    d_{2} &\sim e \cdot R_{a} \sim (1.602 \times 10^{-19}\text{C}) (10^{-10}\text{m}) \\
    &\sim 10^{-29} \text{C} \cdot \text{m}=3\;\text{D} ,    
\end{aligned} \label{d2}
\end{equation}
where $R_{a} \sim 10^{-10}\text{m}$ represents the average atomic radius and $e$ is the electron charge. Here we have used as a unit for the dipole the Debye, which in standard units is $1\;\text{D}= 3.336 \times 10^{-30}\;\text{C} \cdot \text{m}$. 
Another fair assumption would be that there is some water vapour left in the vacuum. Water has a dipole of $6.19\times 10^{-30} \text{C}\cdot\text{m}$\cite{lide2004crc}, which is very close to the value for $d_{2}$ given by Eq.(\ref{d2}).

With this expression for $E_{\text{loc}}$, Eqs.~(\ref{N'alphaEloc}), (\ref{Classius-Mossotti}) and (\ref{E_loc}) give the following value for $d_{1}$:
\begin{equation}
    d_{1}=\frac{3 d_{2}}{2 \pi \epsilon_{r}} \left( \frac{\epsilon_{r}-1} {\epsilon_{r}+2} \right) \sim  10^{-30} \;\text{C} \cdot \text{m},
    \label{d1}
\end{equation}
where we have used the same value for $d_{2}$ used in Eq.(\ref{d2}), i.e. $d_{2} \sim 10^{-29}\text{C} \cdot \text{m}$.

Let us now compute the decoherence rate due to this type of interaction. 
We know that the de Broglie wavelength associated with the environmental ions is $\lambda_{0}={2\pi \hbar}/{p_{0}}$. If we consider the particles to be inside a box of temperature $T$, the average momenta of one ion will be $p_{0} = \sqrt{2mk_{B}T}$, which gives:
$\lambda_{0}={2\pi \hbar}/{\sqrt{2mk_{B}T}} \sim 10^{-9}\text{ m}$, where, we have used $m \sim 10^{-27}\text{ kg}$ and $T \sim 1\;\text{ K}$.
Considering a superposition size of $\Delta x \sim 10^{-5}\text{m}$ (smaller values would give an entanglement that is too small to measure), and a mass for the ion similar to the one of the proton $m \sim 10^{-27}\text{ kg}$, we are in the short-wavelength limit if
\begin{equation}
\begin{split}
     &\lambda_{0}=\frac{2\pi \hbar}{\sqrt{2mk_{B}T}} \ll \Delta x \\
     &\Rightarrow T \gg \frac{4 \pi^2 \hbar^2}{2mk_{B} \Delta x^2} \sim 10^{-7}\text{K},    
\end{split}
\end{equation}
which is always satisfied in a QGEM setup, considering that $T \sim 1\text{K}$. Therefore, the short-wavelength limit will be the right approximation to perform if we want to analyze the QGEM experiment and the decoherence rate will be therefore given by Eq.~(\ref{69}).

We consider the two dipoles $d_{1}$ and $d_{2}$ to have the values as given in Eq.~(\ref{d1}) and Eq.~(\ref{d2}), corresponding to the physical situation where the environmental dipole $d_{2}$ induces a dipole $d_{1}$ inside the crystal because of its dielectric properties. 
Furthermore we take $m \sim 10^{-27}\text{ kg}$ (e.g. Helium molecule), $T \sim 1\;\text{ K}$, $\Delta x \sim 10^{-5}\text{ m}$, $R \sim 10^{-6}\text{ m}$ and $n \sim 10^{8}\text{ m}^{-3}$. This last value can be related also to the pressure $p$, which is the actual parameter that is controlled in a QGEM setup proposal. 
We can consider $n$ and $p$ to be correlated through the perfect gas law $p=n k_{B} T$, which gives the corresponding value for the pressure of the order of $p \sim  10^{-15}\;\text{ Pa}$.

In a QGEM setup, we have that $R \sim 10^{-6}\,\mathrm{m}$ and $\bar{p} \sim 10^{-25}\,\mathrm{kg\,m\,s^{-1}}$, giving ${R\bar p}/{\hbar} \sim 10^3 \gg 1$. This means that, in the final expression for the short-wavelength approximation of the decoherence rate given in Eq.~(\ref{final_dec_long}), the dominant term inside the parenthesis is $32 \left({R\bar p}/{\hbar }\right)^4$, while the others can be considered negligible compared to this one. This gives the following approximation for the decoherence rate:
\begin{equation}
    \Gamma_S \simeq (2 \pi)^{\frac{3}{2}} \frac{2 d_{1} ^2 d_{2}^2 n}{3 \epsilon_{0}^2 \hbar^2 R^2} \sqrt{\frac{2m}{k_{B} T} }.   
    \label{69}
\end{equation}
Filling in all the values for the variables that appear in the final decoherence rate as discussed above, Eq.~(\ref{69}) gives:
\begin{equation}
    \Gamma_{S} \sim 10^{-9}\;\text{Hz}.
    \label{num_value_short}
\end{equation}
%

This means that, if the value for $\Gamma_{S}$ found in Eq.~(\ref{num_value_short}) holds, the off-diagonal elements of $\rho_{S}$ will be suppressed in a time $t_{\text{supp}}$ of the order of $t_{\text{supp}} \sim 10^{9}\text{s}$. This is a very high value, especially compared to the time $\tau \sim 1\;\text{s}$ during which the crystal is kept in superposition in the QGEM proposal. Therefore, assuming that the environmental dipole $d_{2}$ generates a crystal's dipole $d_{1}$ given by Eq.~(\ref{d1}), the dipole-dipole interaction analyzed in this work should not give problems for the realization of the experiment. 
\captionsetup{
    format=plain,
    justification=justified, 
    singlelinecheck=false,
    labelsep=period,
    labelfont=bf,
    textfont=normalfont
}

\begin{figure}
\centering
\includegraphics[scale=0.582]{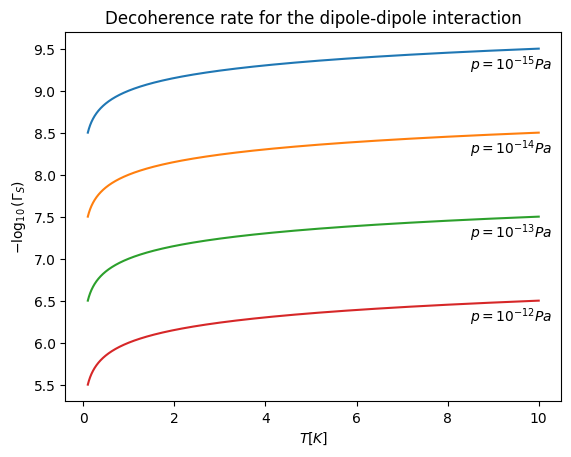}
\caption{This figure illustrates the relationship between the decoherence rate $\Gamma_{S}$, calculated in the short-wavelength limit using Eq.(\ref{69}), and the temperature $T$. In particular, the scenario under investigation corresponds to the scenario where the environmental dipole $d_{2}$ (assumed to be on the order of $d_{2} \sim 1\;\text{D}$ \cite{table}) \textit{induces} an electric dipole within the micro-crystal, $d_{1}$ due to its dielectric properties, as is analyzed in Eq.(\ref{d1}). 
With an environmental particle's mass of $m \sim 10^{-27}\text{kg}$, $T \sim 1\;\text{K}$, $\Delta x \sim 10^{-5}\text{m}$ and the crystal's radius $R \sim 10^{-6}\text{m}$. For a range of temperatures $T \in \{ 10^{-1}\text{K},10\;\text{K} \}$, the figures shows $-\log_{10}({\Gamma_{S}}/{\text{Hz}}) \in \{ 5,9 \}$ which corresponds to a decoherence rate $\Gamma_{S} \in \{ 10^{-9}\text{Hz},10^{-5}\text{Hz} \} $.}
\label{short_n,T}
\end{figure}

\begin{figure}
\centering
\includegraphics[scale=0.575]{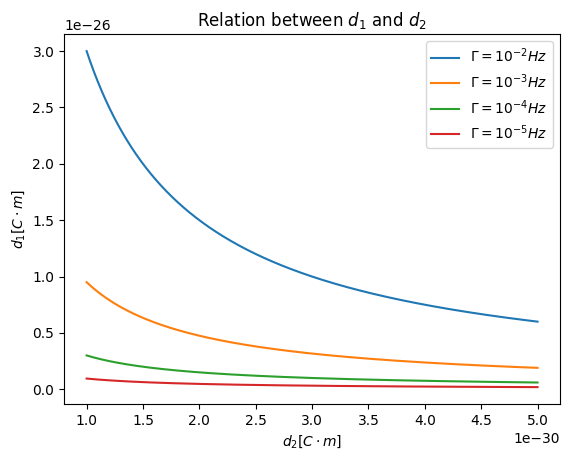}
\caption{This figure is illustrates the relationship between the crystal's permanent dipole $d_{1}$, and the environmental dipole $d_{2}$ for a given decoherence rate $\Gamma_S$. These two variables are related through Eq.(\ref{69}). The various plots with different colours correspond to different values for the decoherence rate $\Gamma_{S} \in \left\{ 10^{-5} \text{H z}, 10^{-2} \text{ Hz} \right\}$. The values for all the other parameters that appear in Eq.(\ref{69}) are $m \sim 10^{-27}\text{kg}$, $T \sim 1\;\text{K}$, $\Delta x \sim 10^{-5}\text{m}$ and $R \sim 10^{-6}\text{m}$. The graph shows that if $d_{2}$ is considered to be of the order of $d_{2} \sim 3.3 \times 10^{-30}\;\text{C} \cdot \text{m}$ \cite{table}, then $d_{1}$ will be of the order of $d_{1} < 10^{-26}\;\text{C} \cdot \text{m} = 10^{-1} \text{e} \cdot \mu \text{m}$ if we require $\Gamma_{S} < 10^{-2}$ Hz.
}
\label{d_1-d_2}
\end{figure}

The general behaviour of the decoherence rate is shown in figure~\ref{short_n,T}. 
Figure \ref{short_n,T} shows the behaviour of $-\log_{10}(\frac{\Gamma_{S}}{\text{Hz}})$ as a function of the ambient temperature $T$, where the different plots correspond to different values for the pressure $p \in \{ 10^{-15}\text{Pa},10^{-12}\text{Pa} \} $ (or, equivalently, for the number density $n$). 
For a range of temperatures $T \in \{ 10^{-1}\text{K},10\;\text{K} \}$, we find $-\log_{10}(\frac{\Gamma_{S}}{\text{Hz}}) \in \{ 5,9 \}$, which corresponds to the decoherence rate in the range: $\Gamma_{S} \in \{ 10^{-9}\text{Hz},10^{-5}\text{Hz} \} $.



\subsection{Constraint for the crystal's dipole} \label{subsec_d1_d2}

The short wavelength limit decoherence rate given in Eq.~(\ref{final_dec_long}) is also used to study the decoherence from the dipole interaction between the permanent dipole of the crystal and the permanent dipole of the environmental particle.
By requiring a maximal decoherence rate we put constraints on the crystal's allowed dipole moment, $d_{1}$. 


Fig.~\ref{d_1-d_2} shows the maximum allowed value of the crystal's dipole, $d_{1}$, as a function of the environmental dipole, $d_{2}$, for different orders of magnitude of the decoherence rate $\Gamma_{S} \in \{ 10^{-5}\text{Hz},10^{-2}\text{ Hz} \} $. 

We require the decoherence rate $\Gamma_{S}$ to be smaller than $10^{-2}\text{ Hz}$, as discussed at the beginning of this section.
The environmental dipole should be associated in a realistic experiment to molecules present, for example, in the air. The values for the electric dipole of this type of molecules center around $d_{2} \sim 1\;\text{D} \simeq 3.336 \times 10^{-30}\text{C} \cdot \text{m}$ \cite{table}, see Eq.~(\ref{d2}). 
With this order for magnitude of $d_{2}$, figure \ref{d_1-d_2} shows that the crystal's dipole $d_{1}$ should be of $d_{1} < 10^{-26}\;\text{C} \cdot \text{ m} = 10^{-1} \text{e} \cdot \mu \text{m}$.

For microspheres with diameters of the order $\sim10\,\mu\text{m}$ the dipole moment has been experimentally estimated to be of the order $\sim 10^{-23} \text{C}\cdot\text{m}$ \cite{Afek:2021bua}. This situation is different from the one considered in this paper, where the radius of the micro-crystal in superposition is considered to be of the order of $\sim1\,\mu\text{m}$, i.e. ten times smaller.
Recent studies \cite{Rivic} on measurement of the scaling of the dipole moment suggest that the electric dipole moment $d$ should scale with the number of atoms $N$ of the crystal. Therefore, considering a micro-crystal with a constant number density $n$ of atoms, the total number of atoms $N$ increases with the volume $N= n V = n ({4 \pi}/{3})R^3$. 
This suggests that the dipole will scale as $d_1 \propto R^3$. Assuming that the measurements done in \cite{Afek:2021bua} can be adapted to the situation considered in this paper by scaling $d$ with the radius to estimate the permanent dipole magnitude for our micro-crystals, we find $d_{1}\sim 10^{-26} \text{C}\cdot\text{m}$, which is exactly the magnitude of the upper bound for the crystal's dipole found in this work, see Fig.~\ref{d_1-d_2}.


\subsection{Induced environmental dipoles}\label{subsec_induced_environment}

\begin{table}[t] 
\begin{center}
\begin{tabular}{|| p{18mm} p{18mm} p{18mm} ||} 
 \hline
 particle & $\alpha '$ (\AA$^3$) & $d_2 \,(\text{C}\cdot\text{m})$ \\ [0.5ex] 
 \hline\hline
 N$_2$ & $1.710$ & $3.425 \times 10^{-35}$ \\ 
 \hline
 O$_2$ & $1.562$ & $3.13 \times 10^{-35}$ \\
 \hline
 Ar & $1.664$ & $3.335 \times 10^{-35}$ \\
 \hline
 CO$_2$ & $2.507$ & $ 5.02 \times 10^{-35}$ \\
 \hline
\end{tabular}
\caption{The polarizability $\alpha '$ for the most common air particles: dinitrogen (N$_2$), carbondioxide (CO$_2$), Argon (Ag) and dioxygen (O$_2$), in CGS units~\cite{pollistx}. Converting in SI units gives the polarizability $\alpha = 4 \pi \epsilon_{0} \alpha '$.
The induced dipole is found from the permanent dipole electric field of the crystal ($E$, see Eq.~\eqref{eq:elect_field_dip}) and the polarizability ($\alpha$), see Eq.~\eqref{induced_63}: $\boldsymbol{d}_2 = \alpha \boldsymbol{E}$.
The magnitude of the induced dipole is given in the third column for a crystal diamond dipole moment $d_1 = 10^{-23} \text{C}\cdot\text{m}$ \cite{Afek:2021bua}.}
\label{table:induced_dipole_air}
\end{center}
\end{table}


\begin{figure}[t]
\centering
\includegraphics[scale=0.51]{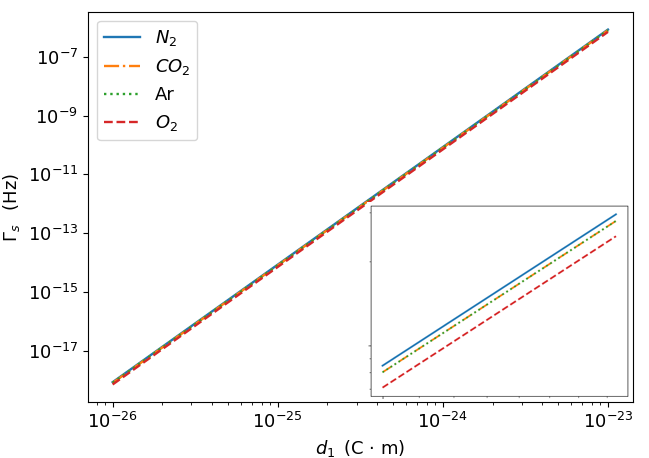}
\caption{This figure shows the decoherence rate $\Gamma_{S}$ from the dipole-dipole interaction between the permanent dipole in the crystal ($d_1$) and the dipole it induces in several types of polarizable environmental particle, as function of the crystal's dipole $d_1$.
The decoherence rate is calculated in the short-wavelength limit, see Eq.(\ref{69}).
The crystal's permanent dipole $d_{1}$ is assumed to be in the range of $d_{1} \in \big\{10^{-26} \text{C} \cdot \text{m}, 10^{-23} \text{C} \cdot \text{m}  \big\} \text{D}$ (see section~\ref{subsec_d1_d2}). 
The polarizability of the most common air compounds: dinitrogen (N$_2$), carbondioxide (CO$_2$), Argon (Ag) and dioxygen (O$_2$), are given in table~\ref{table:induced_dipole_air}. 
Since their polarizabilities are very similar, the lines seem to overlap. In the enlarged figure in the corner one sees that N$_2$ causes the most decoherence, then CO$_2$ and Ar which overlap approximately, and then O$_2$.
With $m \sim 10^{-27}\text{kg}$, $T \sim 1\;\text{K}$, $n \sim 10^{8} 
\text{m}^{-3}$ (or $p \sim 10^{-15} \text{Pa} $), $\Delta x \sim 10^{-5}\text{m}$ and $R \sim 10^{-6}\text{m}$. The graph shows a decoherence rate $\Gamma_{S} \in \{ 10^{-17 }\text{ Hz},10^{-7}\text{ Hz} \} $.}
\label{induced_d2}
\end{figure}


If we assume that the diamond crystal has a measured permanent dipole moment of $d_1 = 10^{-23} \text{C}\cdot\text{m}$ (see section~\ref{subsec_d1_d2},\cite{Afek:2021bua,Rivic}, it could then induce dipoles in the environmental particles as well, resulting in another source of decoherence from the dipole-dipole interaction. To illustrate this, let us consider the content of the air, which consists mostly of nitrogen ($\sim78\%$) and oxygen ($\sim21\%$), with some argon ($\sim 0.9\%$) and carbon-dioxide ($0.04\%$), and very small traces of other molecules \cite{atmosphere}. Additionally one could assume some water vapour is present in the vacuum chamber.
Although, water has a permanent dipole (and is included in the previous discussion of environmental dipoles), the other components do not. However, the electromagnetic field produced by the permanent dipole in the diamond can induce a dipole within the environmental particles, depending on their polarisability.
The induced dipole moment is found from the polarizability and the electric field from the permanent dipole moment of the diamond.
In particular, taking an average distance $\bar{r} = R \sim 10^{-6} \text{m}$ (see section~\ref{subsec_induced_crystal}), the electric field for this average distance is: 
\begin{equation}\label{eq:elect_field_dip}
   E = \frac{d_{1}}{2 \pi \epsilon_{0}} \frac{1}{R^3} 
\end{equation}
which we find to be: $E \sim 1.8 \times 10^{5} {\text{N}}/{\text{C}} $.  The induced dipoles for the most common air particles are given in table~\ref{table:induced_dipole_air}.
Since the taken dipole value $d_1 = 10^{-23} \text{C}\cdot\text{m}$ is for larger spheres then our micro-spheres, this is probably an overestimate since there are indications that dipole moments scale with volume~\cite{Rivic}.

Fig.\ref{induced_d2} shows the behaviour of $\Gamma_{S}$ for different values of the crystal's dipole $d_{1}$. The decoherence rate arises from the dipole-dipole interaction, where the environmental dipole $d_{2}$ is \textit{induced} by the dipole of the crystal, denoted as $d_{1}$. 
For each element presented in table \ref{table:induced_dipole_air}, it is possible to compute the decoherence rate $\Gamma_{S}$ using Eq.~\eqref{69}. It is evident from the figure that when a crystal's dipole of the magnitude $d_{1} \in \big\{10^{-26} \text{C}\cdot\text{m}, 10^{-23} \text{C}\cdot\text{m} \big\}$ induces a dipole moment in the air molecules, the resulting decoherence rate is approximately $\Gamma_{S} \in \big\{10^{-17} \text{Hz}, 10^{-7} \text{Hz} \big\}$. Therefore, the final decoherence time will have a minimum magnitude of the order of $10^{7} \text{s}$, which does not present as a problem in realization the QGEM experiment within $t\sim 1$ s.




\section{Discussion}\label{sec:conclusion}
In this paper, we have analyzed the decoherence rate in the matter-wave interferometer due to the electromagnetic interactions. In particular, we have analyzed 
a special case of dipole-dipole interaction, which plays an important role since generally both the environmental particles (e.g. air molecules) \cite{table,atmosphere,pollistx} and the test mass micro-crystal \cite{Afek:2021bua} can possess an electric dipole moment, or can have an induced dipole from an external electric field. 

In particular, we have used the Born-Markov Master Equation \cite{Schlosshauer:2014pgr,joos2003decoherence}, a well-known tool for investigating the dynamics of quantum systems within a large environment. This model provides a valuable means of deriving a dynamical equation for the density matrix of a quantum system that finds itself immersed in an environment. By employing the Born approximation, which considers the environment to be much larger than the micro-crystal, and incorporating the Markovian assumption, which disregards memory effects within the environment after its interaction with the system, we acquire a non-unitary time evolution equation for the crystal's density matrix. 

In this work, we have delved further into the application of this model, focusing on a generic QED interaction Hamiltonian between two fermions within the framework of the Born-Markov master equation. This Hamiltonian, constructed in terms of destruction/construction operators for both the crystal and the environment, is given by Eq.~(\ref{finalHint}). Using this expression for the QED Hamiltonian interaction inside the Born-Markov master equation, we were able to find Eq.~(\ref{Gamma}), which gives the decoherence rate for the suppression of the off-diagonal elements of the micro-crystal's density matrix. It is worth noting that this final result has the same structure found in the so-called Scattering model, a well know model for decoherence used in the literature \cite{Schlosshauer:2014pgr}.

Further exploring our results, we have directed the focus toward analyzing the dipole-dipole interactions. The first step was to derive the differential cross section for such interactions, as we have done in Eq.~(\ref{diffcsdipdip}). Using this expression for the cross section, we have successfully obtained an explicit expression for the decoherence rate, both in the short-wavelenght limit (see Eq.~(\ref{final_dec_long})) and the long-wavelength limit (see Eq.~(\ref{Gam})).

In section \ref{sec:QGEM}, we have finally applied our results to the QGEM proposal~\cite{Bose:2017nin,vandeKamp:2020rqh,Schut:2021svd}.
In particular, using the parameters employed within the QGEM experiment, such as temperature, pressure, and the dipole of the micro-crystal, we were able to find an explicit expression for the decoherence rate in the short wavelength limit (see Eq.~(\ref{69})). 
The main source of decoherence was found to be from the interaction between the permanent dipole of the crystal and the permanent dipole of the environmental particles (taken to be $d_{2} \sim 1 \; \text{D}$ \cite{table}).
The decoherence from the induced dipoles considered here is negligible compared to the permanent dipole scenario.

Using typical numerical values for the QGEM proposal \cite{Bose:2017nin, Barker:2022mdz}, we were able to find an upper bound for the crystal's dipole, ensuring a decoherence rate smaller than 0.01 Hz. This requirement arises from the necessity of preserving the spatial superposition of the massive particle for a minimum duration of 1 second, accordingly to the QGEM proposal. 
By guaranteeing a decoherence rate of $0.01$ Hz or less, we can confidently ensure the preservation of the crucial superposition throughout the experiment. 
The upper bound for the crystal's dipole determined through this work is on the order of $d_{1} = 10^{-26} \text{C} \cdot \text{m}$, which represents a relatively small value when compared to certain measurements conducted within laboratory settings \cite{Afek:2021bua}.

This research shows that in order to perform the QGEM experiment we need to measure accurately the crystal's permanent dipole for the relevant size and mass.
If the fixed dipole is found to be larger than the upper bound found here, given a certain decoherence rate, then one needs to take measures to mitigate the decoherence due to the dipole moment. Here we do not discuss the strategies to ameliorate this effect but merely point toward a new challenge that the QGEM experiment might have to address these challenges. We will leave that for future investigation.\\


\section*{Acknowledgements} \label{sec:acknowledgements}
MS is supported by the Fundamentals of the Universe research program at the University of Groningen. MT acknowledges funding by the Leverhulme Trust (RPG- 2020-197). S.B. would like to acknowledge EPSRC grants (EP/N031105/1, EP/S000267/1, and EP/X009467/1) and grant ST/W006227/1.
AM’s research is funded by the Netherlands Organisation for Science and Research (NWO) grant number 680-91-119.

\bibliography{dipole.bib} 
\bibliographystyle{ieeetr}

\onecolumngrid
\appendix
\section{Appendix}

\subsection{Differential cross section for a $2 \longrightarrow 2$ process} \label{Appendix}

In this appendix we are going to show the details of some QED calculations. In particular, we are going to find the expression for the matrix element $|\mathcal{M}|^2$ and its relation with the differential cross section $\frac{d \sigma}{d \Omega}$. We will again use the natural units system for simplicity, i.e. $\hbar=c=1$.

Let us start from the definition of cross section \cite{Schwartz:2014sze}:

\begin{equation}
    d\sigma= \frac{1}{T \Phi} dP,
\end{equation}
where $T$ is the total time during which interactions happen, $\Phi$ is the flux of particles (e.g. if we are in the LAB frame then $\Phi$ is the flux of the incoming projectile particles) and $dP$ is the (quantum) probability that one interaction happens.\\
If we consider the COM frame, the flux $\Phi$ will be given by:

\begin{equation}
    \Phi=\frac{|\Vec{v_{1}}-\Vec{v_{2}}|}{V},
\end{equation}

where $\Vec{v_{1}}$ and $\Vec{v_{2}}$ are the velocities of the two initial particles; the minus sign is because they run into each other during the collision.\\
Let us now compute $dP$. During scattering processes, the operator involved is the Scattering matrix $\hat{S}$ and its matrix elements give us the transition probability from an initial state $\ket{i}$ to a final one $\ket{f}$:

\begin{equation}
    dP=\frac{|\bra{f}\hat{S}\ket{i}|^2}{\braket{i} \braket{f}} \prod_{j} \frac{V}{(2\pi)^3}d^3p_{j}.
    \label{probb}
\end{equation}

This formula represents the probability of the interaction inside an infinitesimal volume of the momentum space $d\Pi=\prod_{j} \frac{V}{(2\pi)^3}d^3p_{j}$, where $j$ represents the number of final states.\\
Because in QFT we have $\ket{p}=\sqrt{2E_{p}}\hat{a}_{p}^{\dagger} \ket{0}$, the inner products in the denominator of \eqref{probb} gives:

\begin{equation}
    \braket{i}=\prod_{i}(2\pi)^3 2E_{i} \delta^{(3)}(0).
\end{equation}

In a finite volume, we have:

\begin{equation}
    (2\pi)^3\delta^{(3)}(\Vec{p}=0)=\int_{V} d^3x\; e^{i\Vec{p} \cdot \Vec{x}}=\int_{V} d^3x=V.
\end{equation}

Similarly in $4D$:

\begin{equation}
    \delta^{(4)}(0)=\frac{TV}{(2\pi)^4}.
\end{equation}

This means that:

\begin{equation}
    \begin{cases}
        \braket{i}=(2E_{1}V)(2E_{2}V)\\
        \braket{f}=\prod_{j} 2E_{j}V
    \end{cases}.
\end{equation}

Now, the transferred matrix $\mathcal{T}$ is related to $S$ through:

\begin{equation}
    S=\mathbf{1}+i\mathcal{T}=\mathbf{1}+i(2\pi)^4 \delta^{(4)}(\Sigma p) \mathcal{M}.
\end{equation}

Thus the non-trivial part of the $\hat{S}$ matrix element is:

\begin{equation}
    |\bra{f}\hat{S}\ket{i}|^2=\delta^{(4)}(0)\delta^{(4)}(\Sigma p) (2\pi)^8 |\bra{f}\mathcal{M}\ket{i}|^2=TV\delta^{(4)}(\Sigma p) (2\pi)^4 |\mathcal{M}|^2.
\end{equation}

Plugging everything back in \eqref{probb}, we obtain:

\begin{equation}
    dP=\frac{T}{V}\frac{1}{2E_{1}2E_{2}}|\mathcal{M}|^2 \prod_{j}\frac{d^3p_{j}}{(2\pi)^3 2E_{j}}(2\pi)^4 \delta^{(4)}(\Sigma p).
\end{equation}

Finally, we have an expression for the cross section in the COM:

\begin{equation}
    d\sigma=\frac{1}{2E_{1}2E_{2}|\Vec{v}_{1}-\Vec{v}_{2}|}|\mathcal{M}|^2 d\Pi_{lips},
\end{equation}

where $d\Pi_{lips}=\prod_{j}\frac{d^3p_{j}}{(2\pi)^3 2E_{j}}(2\pi)^4 \delta^{(4)}(\Sigma p)$.\\
In the special case where we have 2 final states, such that $\Vec{p}_{1}=-\Vec{p}_{2}$, $\Vec{p}_{3}=-\Vec{p}_{4}$ and $E_{1}+E_{1}=E_{3}+E_{4}=E_{CM}$, $d\Pi_{lips}$ becomes:

\begin{equation}
    d\Pi_{lips}=\frac{d^3p_{3}}{(2\pi)^3 2E_{3}}\frac{d^3p_{4}}{(2\pi)^3 2E_{4}}(2\pi)^4 \delta^{(4)}(\Sigma p).
\end{equation}

Integrating over $\Vec{p_{4}}$ we obtain:

\begin{equation}
    d\Pi_{lips}=\frac{1}{16\pi^2} d\Omega \int dp_{f} \frac{p_{f}^2}{E_{3}E_{4}}\delta(E_{3}+E_{4}-E_{CM})=\frac{1}{16\pi^2} d\Omega \frac{p_{f}}{E_{CM}} \theta(E_{CM}-m_{3}-m_{4}),
\end{equation}

where $p_{f}=|\Vec{p_{3}}|=|\Vec{p_{4}}|$ and $p_{i}=|\Vec{p_{1}}|=|\Vec{p_{2}}|$.\\
We can now rewrite $|\Vec{v}_{1}-\Vec{v}_{2}|$ as:

\begin{equation}
    |\frac{p_{i}}{E_{1}}+\frac{p_{i}}{E_{2}}|=p_{i}\frac{E_{CM}}{E_{1}E_{2}},
\end{equation}

in order to obtain the final expression for the differential cross section:

\begin{equation}
    \frac{d\sigma}{d\Omega}=\frac{1}{64 \pi^2 E_{CM}^2} \frac{p_{f}}{p_{i}} |\mathcal{M}|^2 \theta(E_{CM}-m_{3}-m_{4}).
    \label{dcs}
\end{equation}

Applying \eqref{dcs} in the case where the target is much heavier than the projectile ($M\gg m$), we can write $  E_{CM} \simeq M$ and also $p_{i}=p_{f}$. In this way, considering the case where $E_{CM} > m_{3}+m_{4}$, we obtain the final expression for the differential cross section:

\begin{equation}
    \frac{d\sigma}{d\Omega}=\frac{1}{64 \pi^2 E_{CM}^2}|\mathcal{M}|^2.
\end{equation}

\subsection{Number density factor} \label{App}

In this Appendix, we will give meaning to the factor $\pi^{-3/2} \Tilde{\sigma}^3$. In particular, let us see the connection between $\Tilde{\sigma}$ and the uncertainty in space along one direction $\sigma_{x}=\sqrt{<x^2>_{\psi}-<x>^2_{\psi}}$ (i.e. along $x$), with $\psi(x)$ being the wave-function in physical space. From \eqref{wfp}, we can see that:

\begin{equation}
\begin{split}
    \psi(x)=\int_{-\infty}^{+\infty} dp_{x} e^{-ip_{x}x} \Tilde{\psi}_{p_{x}}(p)\\
    =\int_{-\infty}^{+\infty} dp_{x} e^{-ip_{x}x} \frac{(2\pi)^{1/2}}{(\pi\Tilde{\sigma}^2)^{1/4}} e^{-\frac{p_{x}^2}{2\Tilde{\sigma}^2}}\\
    =2\pi^{3/4}\Tilde{\sigma}^{1/2}e^{-\frac{\Tilde{\sigma}^2}{2}x^2}.
\end{split}
\end{equation}

Now we can compute $\sigma_{x}$:

\begin{equation}
\begin{split}
   \sigma_{x}=\sqrt{<x^2>_{\psi}-<x>^2_{\psi}}\\
   =\int_{-\infty}^{+\infty} dx \; x^2 |\psi(x)|^2=\frac{\sqrt{2}\pi}{\Tilde{\sigma}}.
\end{split}
\end{equation}

But we know also that the uncertainty in space is physically due to the fact that the environmental particles, as all the other particles involved in the experiment, are confined in a volume $V=L^3$. This means that:

\begin{equation}
    V=L^3=\sigma_{x}^3= \frac{2^{3/2}\pi^3}{\Tilde{\sigma}^3},
\end{equation}

which leads finally to:

\begin{equation}
    \pi^{-3/2}\Tilde{\sigma}^3=\frac{(2\pi)^{3/2}}{V}.
\end{equation}





    \label{ddp}

\end{document}